\documentclass[reprint, superscriptaddress, amsmath, amssymb, aps, pra, floatfix]{revtex4-2}
\usepackage[T1]{fontenc}
\usepackage{graphicx}
\usepackage{dcolumn}
\usepackage{bm}
\usepackage{physics}
\usepackage{amsmath}
\usepackage{xcolor}
\usepackage{hyperref}
\usepackage{cancel}

\begin{document}

\preprint{APS/123-QED}

\title{Nanophotonic cavity cooling of a single atom}
\author{Chenwei Lv}
\affiliation{
 Department of Physics and Astronomy, Purdue University, West Lafayette, IN 47907, USA
}
\author{Ming Zhu}
\affiliation{
 Department of Physics and Astronomy, Purdue University, West Lafayette, IN 47907, USA
}
\author{Sambit Banerjee}
\affiliation{
 Department of Physics and Astronomy, Purdue University, West Lafayette, IN 47907, USA
}
\author{Chen-Lung Hung}
\email{Email: clhung@purdue.edu}
\affiliation{
 Department of Physics and Astronomy, Purdue University, West Lafayette, IN 47907, USA
}
\affiliation{
 Purdue Quantum Science and Engineering Institute, Purdue University, West Lafayette, IN 47907, USA
}

\date{\today}

\begin{abstract}
We investigate external and internal dynamics of a two-level atom strongly coupled to a weakly pumped nanophotonic cavity. 
We calculate the dipole force, friction force, and stochastic force due to the cavity pump field, and show that a three-dimensional cooling region exists near the surface of a cavity. 
Using a two-color evanescent field trap as an example, we perform three-dimensional Monte-Carlo simulations to demonstrate efficient loading of single atoms into a trap by momentum diffusion, and the stability of cavity cooling near the trap center. 
Our analyses show that cavity cooling can be a promising method for directly loading cold atoms from free-space into a surface micro-trap.
We further discuss the impact of pump intensity on atom trapping and loading efficiency.
\end{abstract}

\maketitle

\section{Introduction}

Strong and efficient atom-light interaction has been realized on various nanophotonic platforms~\cite{chang2018colloquium}, such as optical nanofibers~\cite{vetsch2010optical,goban2012demonstration,beguin2014generation,corzo2019waveguide,rajasree2020generation,hummer2021probing,rakonjac2022storage}, nanofiber cavities \cite{kato2015strong,2019PRL_nanoFiberCavity_Nayak}, microring resonators \cite{zhou2023coupling}, photonic crystal waveguides and cavities \cite{thompson2013coupling,goban2014atom,dhordjevic2021entanglement}. 
To further enable quantum control and manipulation with single atoms coupled to these platforms, efficient atom trapping near nanoscale dielectrics is a key requirement. 
Two-color evanescent field traps with far-off resonant red- and blue-detuned lights have been implemented on optical nanofibers~\cite{vetsch2010optical,goban2012demonstration,beguin2014generation,corzo2019waveguide,kato2015strong,le2004atom, vetsch2010optical,ton2022state}. 
Similar proposals were put forward on strip, ridge, or rib waveguide platforms~\cite{ovchinnikov2020towards, chang2019microring} and photonic crystal waveguide~\cite{bouscal2023systematic}. 
More exotic traps based on optical and vacuum force have also been proposed \cite{hung2013trapped,chang2014trapping,gonzalez2015subwavelength,fuchs2018nonadditivity}. 
In these platforms, the optical trap center is typically designed to be $\lesssim 200~$nm from a dielectric surface  to ensure strong atom-light coupling with a large cooperativity parameter $\gtrsim O(10)$. 
The potential depth is typically $\sim O(1)~$mK, and varies rapidly within one micron above the dielectric surface. 
This would demand a highly efficient cooling scheme to slow single atoms into these surface traps in a short traveling distance. 
So far, atom loading using conventional magneto-optical traps (MOT) and polarization-gradient cooling has only been demonstrated on suspended waveguide structures, but not on general photonic platforms. 
Effects like unbalanced radiation pressure from large surface scattered light, limited capture angle facing free-space, and the lack of optical access could all account for the inefficient cooling and loading efficiency.

Here, we propose an optical cooling method that can efficiently stop and load single atoms from free-space onto a nanoscale photonic cavity using a guided mode field. 
Our scheme is based on cavity cooling, with strong atom-light interaction coupling both the \textit{transverse} and \textit{axial} atomic motion to a cavity field in a high-Q nanophotonic cavity. 
We note that the very concept of cavity cooling has been discussed over two decades ago \cite{horak1997cavity,hechenblaikner1998cooling,vuletic2000laser,maunz2004cavity,wolke2012cavity,ritsch2013cold,hosseini2017cavity} with a vast literature focusing on motional coupling of cold atoms in cavities bounded by free-space mirrors. 
Experimentally, it has been demonstrated that single atoms can be trapped inside a high-finesse Fabry-Perot cavity with single photons~\cite{hood2000atom, pinkse2000trapping} and can be further cooled down using a weak cavity probe field~\cite{maunz2004cavity,wolke2012cavity}. 
The intra-cavity cooling mechanism can be understood as a Sisyphus-type cooling in the picture of cavity dressed-states~\cite{horak1997cavity, hechenblaikner1998cooling}, with an atom moving along the cavity axis and with position-dependent coupling to a standing-wave mode. 
Alternatively, cavity cooling can be achieved by Doppler cooling based on preferential scattering into the cavity mode~\cite{vuletic2000laser}. 
We note that these established methods primarily focus on \textit{axial} cooling effects, because significant cooling force arises only when an atom experiences large mode intensity variation in sub-micron distance scales. 
In addition to using a weak cavity probe, free-space cooling beams could also be sent from the side of a cavity~\cite{vuletic2001three, hosseini2017cavity}, inducing two-photon Doppler cooling in two- and three-dimensions (3D). 

 In the context of nanophotonic cavity cooling, one may utilize strong evanescent field gradient surrounding a nanoscale photonic structure to provide a large stopping force. 
 Evanescent field is one defining feature for guided modes in nanoscale waveguides and cavities, as strong dielectric confinement and total internal reflection in a nano-structure make the mode field intensity decay rapidly outside the dielectric boundaries. 
 When an atom approaches the evanescent region from free-space, it could experience a significant transverse cooling effect. 
 The question is whether this cooling force is sufficient to reduce the kinetic energy of a moving atom in a short distance, making the atom trappable. 
 Ref.~\cite{le2010motion} discussed semiclassical dynamics of atomic motion around a weakly-driven nanofiber cavity. 
 Ref.~\cite{Domokos2001} discussed transverse cooling in a bichromatic evanescent trap in the dispersive regime. 
 In Ref.~\cite{Rauschenbeutel2018}, degenerate Raman cooling close to the atom's motional ground state is achieved near the surface of an optical nanofiber.
 To our knowledge, however, transverse cooling in an evanescent field near a weakly-driven nanophotonic cavity has not been investigated systematically.

In this paper, we discuss how a weak cavity pump field, blue-detuned to both the cavity and atomic resonances, can be utilized for atom cooling and surface-trap loading; see Fig.~\ref{fig: atom-cavity model}. 
Specifically, when an atom flies towards the strong-coupling region above a dielectric surface, increased atom-light coupling reduces the intra-cavity photon number, inducing friction.
We show that cavity photon fluctuations could provide a stochastic stopping force for loading a single atom, in a single pass, into an optical trap near the surface. 
In principle, this method could be applied to surface microtraps formed on general nanophotonic cavities.
To show a concrete example, we discuss the cooling effect on a cavity formed by a rectangular waveguide, and introduce a far-off resonant, two-color evanescent field trap to discuss cooling and atom loading efficiency with a variable cavity pump field. 
By selecting the pump frequency detuning and the position of the two-color trap center, a stochastically loaded atom would continue to experience damping force and small momentum diffusion, leading to a low equilibrium temperature similar to the case discussed in conventional cavity cooling~\cite{van2001cooling}. 

The paper is organized as follows. 
In Sec.~\ref{sec: model}, we review the semiclassical model of an atom coupled to a driven cavity mode. 
We then derive the analytical expressions of the dipole force, friction, and diffusion coefficients in the weak-driving limit, validate and extend the results to large pump rates with full quantum solutions by numerically solving the Lindblad master equation. 
In Sec.~\ref{sec: cooling}, we investigate the cooling mechanism and estimate the equilibrium temperature by cavity cooling together with a two-color evanescent field trap. 
In Sec.~\ref{sec: MC simulation}, we apply 3D Monte-Carlo simulations to verify the stability of atom trapping and determine the atom loading rate. 
The influence of cavity probe intensity on cooling and trapping is then discussed.

\section{\label{sec: model}Semiclassical Model of an atom near a nanophotonic waveguide cavity}

\begin{figure}[t]
    \centering
    \includegraphics[width=\columnwidth]{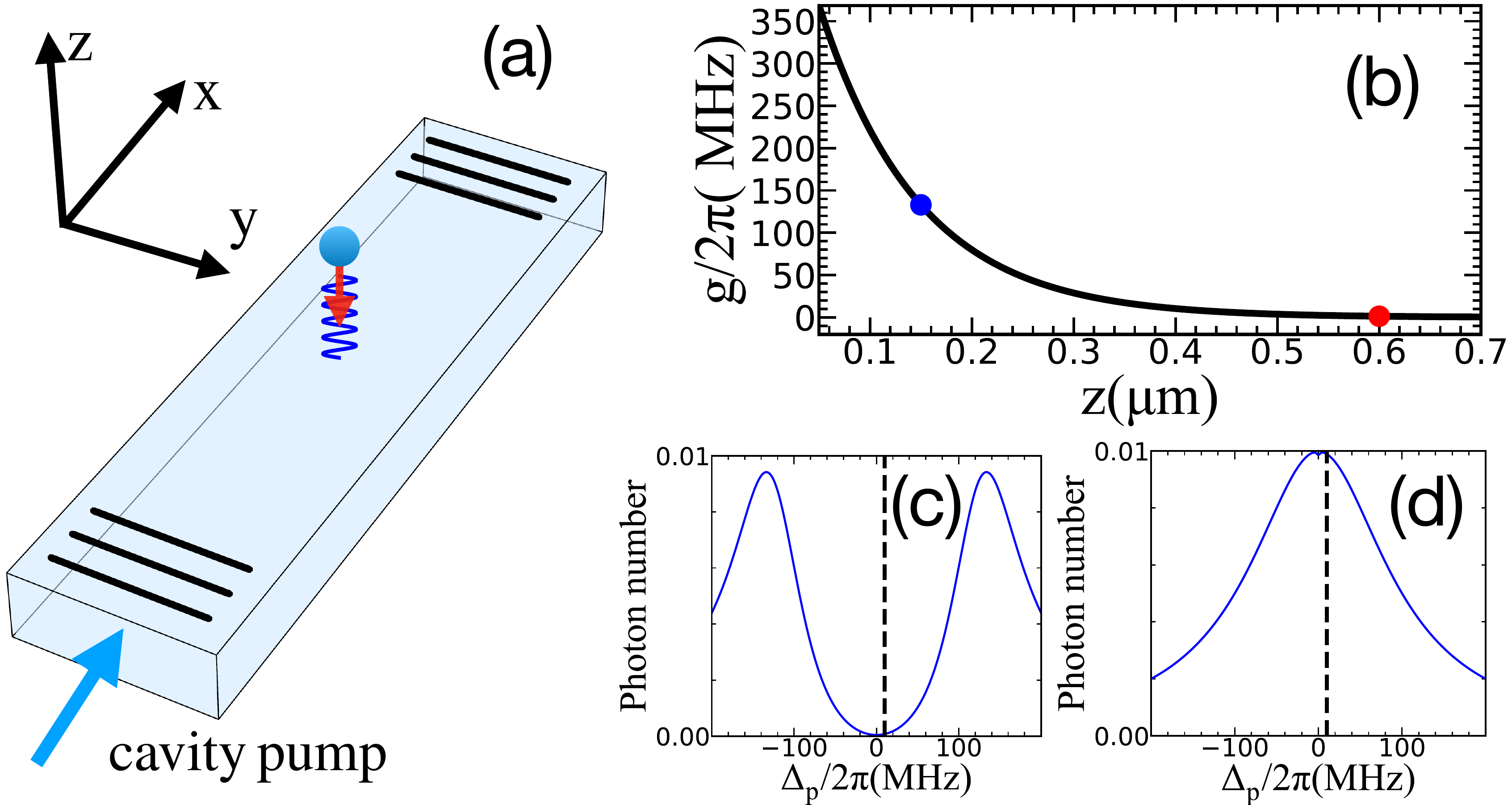}
    \caption{Nanophotonic cavity cooling of single atoms. 
    (a) Cavity cooling for a single atom (blue circle) approaching the strong coupling region near a nanophotonic cavity driven by a blue-detuned pump. 
    (b) Atom-photon coupling strength $g(z)$ at the waveguide center ($y=0$) and at an anti-node of the cavity mode ($x=0$). 
    Top surface of the waveguide is at $z = 0$. 
    (c-d) Cavity photon number for an atom in the strong-coupling (c) and weak-coupling regimes (d), marked by blue and red circles in (b), respectively, as a function of pump detuning $\Delta_p$ with a pump rate of $\varepsilon=2\pi\times 10$ MHz. Cavity resonance is aligned with free-space atomic resonance. 
    }
    \label{fig: atom-cavity model}
\end{figure}

We first describe the system and establish a model of a two-level atom interacting with a quantized cavity mode. 
As shown in Fig.~\ref{fig: atom-cavity model}(a), a laser-cooled atom is released from a MOT and approaches a nanophotonic cavity by time-of-flight or by optical guiding \cite{xu2001efficient,zhou2023coupling}. 
We consider an optical cavity along a simple rectangular waveguide. 
The cavity is bounded by reflective elements such as photonic crystal mirrors or Bragg gratings. 
The cavity mode field forms a standing wave, and can be excited by pump light from one end of the waveguide. 

Considering only the internal degrees of freedom, this atom-cavity system could be described by the Jaynes-Cumming Hamiltonian, 
\begin{equation}
\label{eq: H_JC}
\hat{H}_{\rm JC} =-\hbar \Delta _{\text{c}} \hat{a}^{\dagger } \hat{a} -\hbar \Delta_{\text{a}} \hat{\sigma}_{+}\hat{\sigma}_{-} + \hbar g({\bf x}) \left( \hat{a}^{\dag} \hat{\sigma}_{-} + \hat{a} \hat{\sigma}_{+} \right)
\end{equation}
where $\hat a^\dag (\hat a)$ is the creation (annihilation) operator of the cavity mode, $\hat\sigma_+=\ket{e}\bra{g}$ couples the atomic ground state $\ket{g}$ to the excited state $\ket{e}$, $\hat\sigma_-=\hat\sigma_+^\dag$, $\Delta_{\text{c}(\text{a})} = \omega_\mathrm{l}-\omega_\mathrm{c (a)}$ is the detuning of pump frequency $\omega_\mathrm{l}$ from the cavity (atomic) resonance $\omega_\mathrm{c(a)}$, and $\hbar$ is the reduced Planck constant.
The coupling strength $g({\bf x}) = \sqrt{3 \pi\gamma c^3/2 V_{m}({\bf x})\omega_\text{a}^{2}}$ depends on the mode volume $V_{m}\equiv \int \epsilon ({\bf x}')|\mathbf{E}({\bf x}')|^2 d^3x'/\epsilon ({\bf x})|\mathbf{E}({\bf x})|^2$ at the atom location ${\bf x}=(x,y,z)$~\cite{chang2019microring}, where $\epsilon ({\bf x})$ is the dielectric function, $\gamma$ is the atomic decay rate in free-space, and $c$ is the speed of light. 
In the evanescent field region outside a rectangular waveguide, the mode field strength (a standing-wave mode) can be approximated by the functional form
$|\textbf{E}({\bf x})| \propto \cos(k_\mathrm{ax} x ) \cos(y/q) \exp(-z/d)$,
where $k_\mathrm{ax}$ is the axial wavenumber, $q$ and $d$ are two constant lengths. 
Near the mid-plane of the waveguide ($y=0$) and close to an anti-node of the cavity mode, the atom-photon coupling strength can be written in the approximate form
\begin{equation}
\label{eq: g}
g({\bf x}) = g_{0} \cos(k_\mathrm{ax} x ) \cos(y/q)\exp(-z/d)\,,
\end{equation}
where $g_{0}=g(0,0,0)$ is the maximum coupling strength on the waveguide surface. 
Figure~\ref{fig: atom-cavity model}(b) displays a sample atom-photon coupling strength, calculated by applying an approximate analytical description for the fundamental TM-like mode~\cite{Westerveld2012} with waveguide parameters as listed in Table~\ref{tab: MC_parameters} and for a pump field at a free-space wavelength $\lambda = 852$~nm close to the D2 line resonance of atomic cesium.
The following discussions all use cesium as an example.  

Exciting the cavity via an external pump at a rate $\varepsilon$, the dynamics is described by 
\begin{equation}
\label{eq: H_pump}
\hat{H}_{\rm pump} = -i \hbar  \left(\varepsilon \hat{a}^{\dagger} -\varepsilon ^{*} \hat{a}\right).
\end{equation}
Figure~\ref{fig: atom-cavity model}(c) [(d)] shows the cavity photon number versus pump detuning with an atom closer to (far away from) the waveguide in the strong (weak) coupling region, $g(z) \gtrsim \kappa$ [$g(z)\lesssim \kappa$], where we consider a simple case with the cavity resonance aligned to the atomic transition frequency in free-space. 
The pump detuning is denoted as $\Delta_\text{p} = \Delta_\text{c} = \Delta_\text{a}$. 
As we will discuss in the following, this configuration results in transverse cooling when an atom approaches the waveguide under a proper detuning $g(z_\text{t})>\Delta_\mathrm{p}>0$, where $z_\text{t}$ is a designated trap center.

We note that, in the presence of an optical trap near the surface, the atomic resonance can be shifted due to the differential AC Stark shift between the ground and the excited states. 
This may introduce transient heating effect when an atom is in the excited state, and can also complicate our cooling analyses. 
However, one may adopt magic wavelengths for atomic species such as Cs~\cite{katori2003ultrastable,chang2019microring}, Sr \cite{ido2003recoil, ton2022state}, and Yb\cite{guo2010dipole} to cancel the differential light shifts.  
For simplicity of discussions, in the following, we neglect the contribution of differential light shifts induced by an optical trap near the surface. 

We now consider the full dynamics of the atom-coupled system. By taking into account of resonator photon decay rate $\kappa$ and the atomic decay rate ($\Gamma=\gamma/2$), the Lindblad master equation of the coupled system is written as
\begin{equation}
\begin{aligned}
    \frac{\partial\hat\rho}{\partial t} = \mathcal{L}\hat\rho = & -\frac{i}{\hbar} \left[ \hat{H}_{\rm JC}+\hat{H}_{\rm pump} , \hat\rho \right] + \kappa\mathcal{D}[\hat{a}]\hat\rho +\Gamma \mathcal{D}[\hat{\sigma}_{-}]\hat\rho,
\end{aligned}
\label{eq: master equation}
\end{equation}
where $\hat\rho$ is the density matrix of the composite atom-photon system, $\mathcal{L}$ is the Liouvillian superoperator and the dissipator takes the form $\mathcal{D}[\hat{b}] \hat\rho=2 \hat{b} \hat\rho \hat{b}^{\dagger}- \hat{b}^{\dagger} \hat{b} \hat\rho - \hat\rho \hat{b}^{\dagger} \hat{b}$.

While we apply a quantum treatment for the system's internal degrees of freedom, the atomic motion is assumed to be moving much slower than the cavity dynamics. 
This is justified since, with initial laser cooling, the atomic temperature is below the Doppler limit, and the starting velocity $v$ satisfies the requirement $k v \ll \Gamma, \kappa$, where $k$ is the photon wavenumber. 
We thus make an approximation that the atom-photon dynamics is in the steady-state at any instantaneous time, and the atomic motion is treated semiclassically by a stochastic equation~\cite{Dalibard1985,van2001cooling}, 
\begin{equation}
    \begin{split}
       M \frac{d^2x^i(t)}{dt^2} =& -\partial_i U({\bf x})+\langle \hat F^i({\bf x})\rangle - \beta^{ij}({\bf x})v_j(t)\\
        &+B^{ij}({\bf {x}})w_j(t)\,,
    \end{split}
\end{equation}
where $i, j =x,y,z$ labels the Cartesian coordinates. The repeated indices follow a summation convention (We do not distinguish between upper and lower indices). 
Here $M$ is the atomic mass, and $U({\bf x})$ is the sum of the surface trap potential and the atom-surface Casimir-Polder interaction~\cite{le2010motion}. 
The pump-field induced steady-state dipole force $\langle \hat F^i\rangle$, the friction tensor $\beta^{ij}$, and the diffusion tensor ${\bf D}={\bf BB}^T/2$ will be determined after we evaluate the steady-state from the Lindblad master equation in the following sections.
To describe momentum diffusion, $w_j(t)$ denotes a Gaussian random variable with zero mean and unit variance \footnote{We note that while three random variables are needed to describe 3D momentum diffusion caused by, e.g., spontaneous emission, one random variable suffices to describe dipole force amplitude fluctuations since the force orientation is always along $\nabla g$. This quoted equation works for general cases nonetheless.}. Beyond this semiclassical model, a complete quantum treatment considering quantized atomic motion in a cavity QED field has been discussed in Ref.~\cite{walther1998quantum}.

\subsection{Steady-state force on a motionless atom}

For a stationary atom, the Liouvillian superoperator is a constant in time. 
The steady-state force reads
\begin{equation}
\label{eq: force 0}
\begin{aligned}
\expval{\hat {\bf F}({\bf x})} 
= & -\expval{\nabla \left( \hat{H}_{\rm JC}+\hat{H}_{\rm pump}\right)} \\ 
= & - \hbar \nabla g({\bf x}) \expval{\hat{a}^{\dag} \hat{\sigma}_{-} + \hat{a} \hat{\sigma}_{+}},
\end{aligned}
\end{equation}
where $\langle\hat{O}\rangle = {\rm Tr}(\hat{O} \hat{\rho}_0)$.
Here, $\hat{\rho}_0$ represents the density matrix in the limit of a stationary atom at ${\bf x}(t)={\bf x}_0$. 
We solve for the steady-state density matrix using the equation $\mathcal{L}({\bf x_0})\hat{\rho}_0=0$. 

Using the master equation, we first analytically calculate the steady-state density matrix in the weak-driving limit, where both the excited state population and the photon number are small. 
We can truncate the Hilbert space of the atom-photon system into $\ket{1,g},\ket{0,e}$ and $\ket{0,g}$, where $0,1$ denotes the photon number. 
To the leading order of the pumping strength $\varepsilon$, we obtain
\begin{equation}
 \begin{split}
     &\hat \rho^{\rm weak}_0=\frac{1}{|Q|^2} \times
     \\&\begin{pmatrix}
    (\Gamma^2+\Delta_a^2)\varepsilon^2&g(\Delta_a+i\Gamma)\varepsilon^2&Q^*(\Gamma-i\Delta_a)\varepsilon\\
    g(\Delta_a-i\Gamma)\varepsilon^2& g^2\varepsilon^2&-iQ^*g\varepsilon\\
    Q(\Gamma+i\Delta_a)\varepsilon&iQg\varepsilon& \chi
  \end{pmatrix}
 \end{split}\label{eq:density_weak}\,,
\end{equation}
where $\chi=|Q|^{2}-(\Gamma^2+\Delta_a^2+g^2)\varepsilon^2$ such that $\Tr(\hat{\rho}^{\rm weak}_0)=1$, and $Q=\Gamma \kappa+g^2-\Delta_a\Delta_c-i(\Delta_c\Gamma+\Delta_a\kappa)$. 
The steady-state force in the weak-driving limit is thus
\begin{equation}
    \expval{\hat{\bf F}({\bf x})} 
    =-\frac{\varepsilon^2}{|Q({\bf x})|^2}\hbar(\nabla g^2({\bf x}))\Delta_\text{a} \,.
\end{equation}
Similarly, we can write down the average cavity photon number
$\Bar{N}({\bf x}) = \expval{\hat{a}^{\dagger} \hat{a}}  = \frac{\varepsilon^{2}}{|Q({\bf x})|^2} \left( \Delta_{a}^{2} + \Gamma^{2} \right)$, 
and the excited state population 
$P_{e}({\bf x}) = \expval{\hat{\sigma}_{+} \hat{\sigma}_{-}}  =  \frac{\varepsilon^{2}}{|Q({\bf x})|^2} g({\bf x})^{2}$.

\begin{figure}[t]
    \centering
    \includegraphics[width=1 \columnwidth]{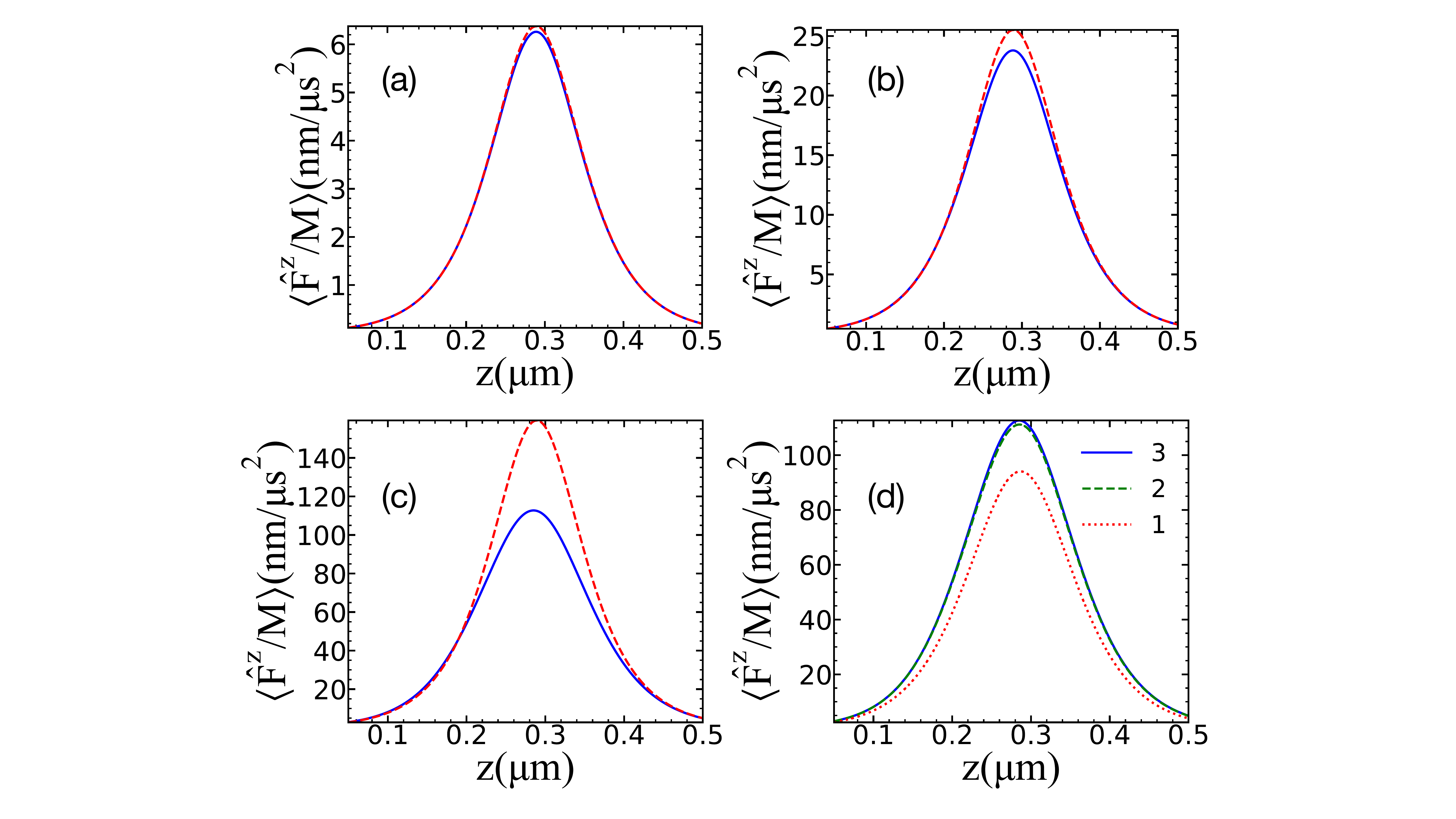}
    \caption{
    Steady-state force $\langle \hat{F}^z(0,0,z)\rangle$ versus atom position $z$, calculated using $\varepsilon/2\pi =$ (a) 5, (b) 10, and (c, d) 25~MHz, respectively, 
    together with system parameters as listed in Table~\ref{tab: MC_parameters} and $g(z)$ as shown in Fig.~\ref{fig: atom-cavity model}(b). 
    Solid (dashed) curves are numerical (analytical) results. 
    In numerical calculations, the Hilbert space is truncated at $N=4$ photons in (a-c) and $N=1,2,3$ as labeled in (d).
    }
    \label{fig: cavity pump potential comparison}
\end{figure}

Since we are interested in the \textit{transverse} motion of an atom approaching the waveguide, we first discuss the magnitude of the most relevant force and coefficients, which are along the $z$-axis, and show their values on the mid-plane of a waveguide at $y=0$.
We will discuss the effect of other directions in a later section.
To provide a concrete example, we assume a cavity decay rate $\kappa = 2\pi \times 100~$MHz, corresponding to a quality factor $\bar Q=\omega_\text{c}/(2\kappa) \approx 1.5\times 10^6$ for state-of-the-art nanophotonic cavities. 
For reasons we will discuss later, the pump detuning is chosen as $\Delta_{\rm p}  = 2\pi \times 10~$MHz, and the atomic decay rate is $\Gamma = 2\pi \times 2.61~$MHz (for Cs D2 line), as summarized in Table.~\ref{tab: MC_parameters}. 

\begin{table}[b]
 \caption{\label{tab: MC_parameters} Parameters used for simulation}
\begin{ruledtabular}
\begin{tabular}{lcc}
\textrm{Parameter}&
\textrm{Symbol}&
\textrm{Value}\\
\colrule
Length of waveguide & L & 164$\mu$m\\
Width of waveguide  & W & 950nm\\
Height of waveguide & H & 360nm\\
Atomic spontaneous decay rate & $\Gamma$ & $2 \pi \times 2.61\mathrm{MHz}$ \\
\colrule
Cavity decay rate          & $\kappa$  & $2 \pi \times 100\mathrm{MHz}$ \\
Cavity pump laser detuning & $\Delta_{p}$ & $2 \pi \times 10\mathrm{MHz}$\\
\end{tabular}
\end{ruledtabular}
\end{table}

To validate the weak-driving approximation, we numerically evaluate the steady state $\hat{\rho}_0$ and calculate the expectation value $\langle\hat{\mathbf{F}}\rangle$.
With a pump rate of $\varepsilon =2\pi \times  5~$MHz, in Fig.~\ref{fig: cavity pump potential comparison}(a), we show the agreement between the numerical and analytical results of $\langle\hat{F}_{z}\rangle$. 
The force points away from the waveguide, effectively forming a repulsive potential. 
This potential barrier can be easily overridden by a two-color trap that we will introduce later.
Its fluctuations, as we will discuss next, lead to momentum diffusion.

As the pump rate increases, simulated force is smaller than the analytical result due to saturation of the excited state population. 
The difference between the predicted maximum force is 10\% with $\varepsilon =2\pi \times  10~$MHz in Fig.~\ref{fig: cavity pump potential comparison}(b) and 50\% with $\varepsilon =2\pi \times  25~$MHz in Fig.~\ref{fig: cavity pump potential comparison}(c). 
In the numerical calculations, we also consider different truncated photon numbers in the Hilbert space. 
Figure~\ref{fig: cavity pump potential comparison}(d) shows that the maximum cavity pump force approaches an upper limit as the truncated cavity photon number goes beyond $2$, indicating that primarily $N\leq3$ Fock states are occupied at the largest considered pump rate ($\varepsilon=2\pi\times25~$MHz). 
In the following numerical simulations, we truncate the Hilbert space to photon number $N=4$ as a trade-off between result accuracy and computation resources.

Lastly, we comment that the peak position of $\langle\hat F^z\rangle$ occurs at $g(z)=[(\Delta_\text{p}^2+\Gamma^2)(\Delta_\text{p}^2+\kappa^2)]^{1/4}>|\Delta_\text{p}|$ for weak-driving. 
Regardless of the magnitude of the pump detuning, this position always occurs when the pump becomes red-detuned (blue-detuned) to the frequency of the upper (lower) cavity dressed-state.

\subsection{Friction force on a slowly moving atom}

We now review the system dynamics with a slowly moving atom. 
As a leading order correction to the stationary case, we replace the $\partial_t$ in the master equation Eq.~(\ref{eq: master equation}) with the hydrodynamic derivative $\partial_t+{\bf v}\cdot\nabla$ and consider an expansion of the density matrix with respect to the velocity. 
Inserting $\hat\rho = \hat\rho_{0} + \hat\rho_{1}^i v_i + ...$ into
\begin{equation}
    \left( \frac{\partial}{\partial t} + {\bf v} \cdot \nabla \right) \hat\rho = \mathcal{L}\hat\rho\,,
\end{equation}
and, to the first order of ${\bf v}$, we find \cite{van2001cooling, hechenblaikner1998cooling, le2010motion} 
\begin{equation}
 v_i \partial_i \hat\rho_{0} = \mathcal{L}\hat\rho^i_{1}.
\end{equation}
The leading order correction to the steady-state force is
\begin{equation}
 F^i_{1}= -{\rm Tr}\left(v_j\hat\rho^j_{1} \partial_i \left( \hat{H}_{\rm JC}+\hat{H}_{\rm pump}\right) \right) = \beta^{ij}v_j,
\end{equation}
where $\boldsymbol{\beta}$ is a $3\times 3$ tensor. 
When $\beta^{ii} < 0$, the friction force is opposite to the velocity, serving as a damping force along the $i$-direction. 
With $\beta^{ii} > 0$, on the contrary, friction force heats the atom. 
The off-diagonal terms lead to velocity transfer between different directions. 
In the weak-driving limit, we can calculate $\hat\rho^{i,\rm weak}_{1}$ using $\hat\rho_{0}^{\rm weak}$, such that an analytical formula could be derived. 
Due to the complexity of expression, we show the formula in Appendix~\ref{app: friction}. 

\begin{figure}[b]
    \centering
    \includegraphics[width=1 \columnwidth]{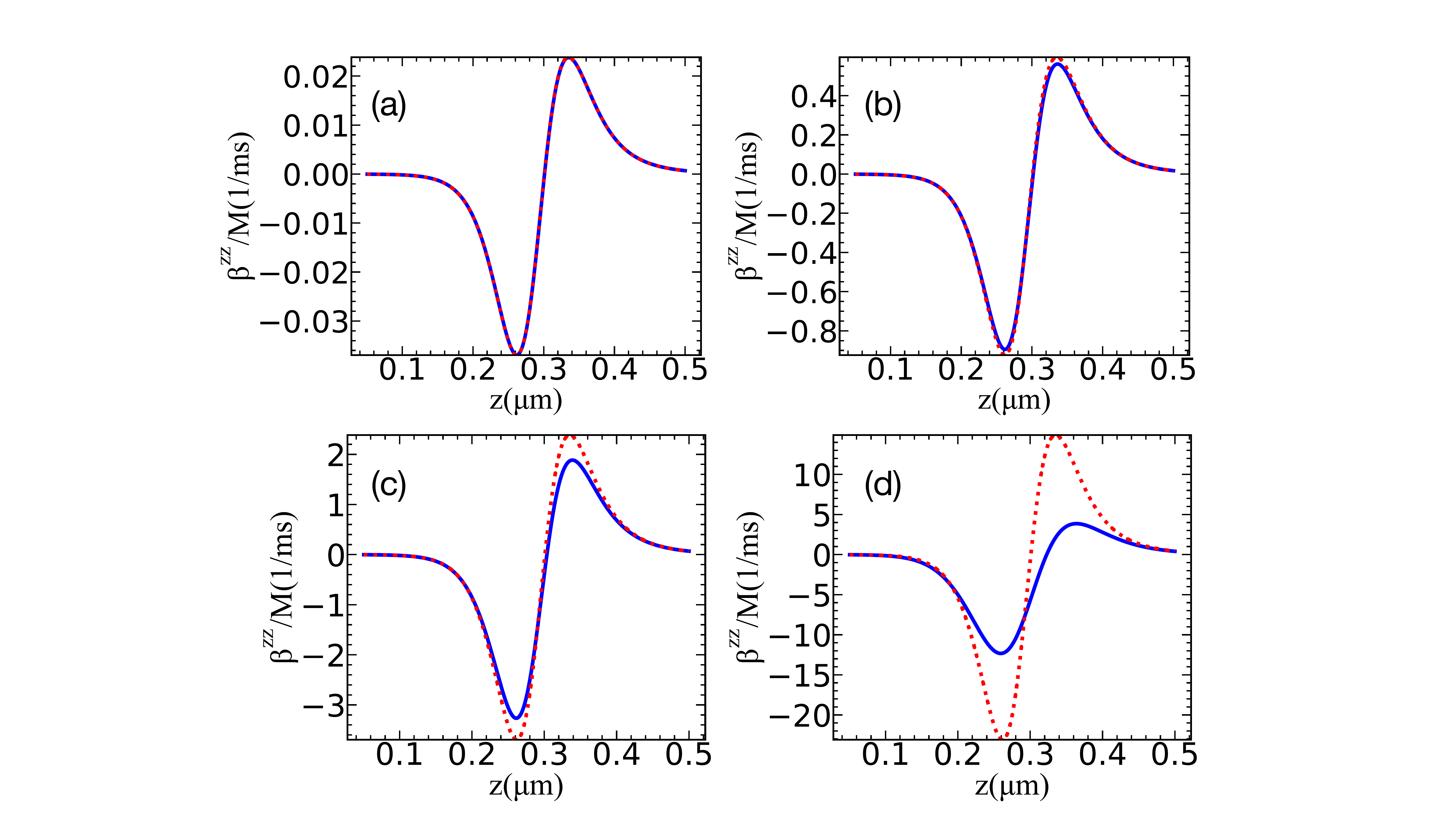}
    \caption{
    Position-dependent friction coefficients $\beta^{zz}(0,0,z)$ calculated with $\varepsilon/2\pi = $ (a) 1, (b) 5, (c) 10, (b) 25 MHz, respectively. 
    Solid curves (dashed curves) are numerical (analytical) results. 
    }
    \label{fig: friction comparison}
\end{figure}

Figure~\ref{fig: friction comparison} displays the friction coefficient $\beta^{zz}$ along the $z$-direction.
In the weak-driving limit, numerical and analytical results match well, as expected. 
At a stronger pump rate, saturation effect again leads to weaker $\beta^{zz}$ compared to the weak-driving approximation.
In addition, the position of the maximum damping force moves slightly closer to the waveguide as the cavity pump intensity increases. 

We comment that friction can cause heating when an atom is away from the waveguide, but cools when an atom moves across the region where $\langle\hat{F}^z\rangle$ maximizes and the pump becomes red-detuned to the cavity dressed-state, where a stronger atom-photon coupling reduces the cavity photon number. 
The origin of damping is explained by a Sisyphus-type cooling~\cite{horak1997cavity, hechenblaikner1998cooling}, considering the delayed response of intra-cavity photon number with respect to the atomic motion. 
As an atom moves closer to the waveguide with increasing $g$, a delayed reduction of intra-cavity pump photons (due to finite atom-cavity interaction time) causes an atom losing more kinetic energy from the repulsive force, and this results in damping. 
We note that cooling can also occur with a red-detuned pump $\Delta_{\rm p}<0$, although the cooling zone is located farther away from the waveguide and $\Delta_{\rm p} < -g$.  

Considering the magnitude of the friction coefficient in Fig.~\ref{fig: friction comparison}, it requires a millisecond time-scale, even for stronger pump intensities, for the cooling/heating effect to become significant. 
As such, while friction may keep a trapped atom cold, this effect alone is \textit{insufficient} to load a free-space atom into a surface trap, which typically traverses the trap region in a microsecond times scale. 
In the next section, we discuss the magnitude of momentum diffusion and explain why it provides sufficient stochastic stopping force to load an atom.

\subsection{Steady-state diffusion coefficient}
In previous sections, we discuss the dipole force of a motionless state and the possible cooling friction force of a moving atom on specific conditions. 

The atom velocity will reduce to zero if it stays in the damping region.
However, atomic momentum spread prevents the motion from reaching the ground state and leads to a finite equilibrium temperature for a stably trapped atom. 
This effect is characterized by the diffusion coefficient. 
Moreover, as we will show, momentum diffusion is also an important mechanism for providing a large enough stochastic stopping force to reduce the kinetic energy of an atom upon entering the trap region.

The fluctuation of force results in the diffusion tensor of the form ~\cite{gordon1980motion, Dalibard1985}
\begin{equation}
    D^{ij} = {\rm Re}\int^{\infty}_0 dt \left[ \expval{\hat F^i(0)\hat F^j(t)} - \expval{\hat F^i(0)} \expval{\hat F^j(t)} \right],
\label{eq: diffusion definition}
\end{equation}
where $\hat F^i(0)=-\hbar\partial_i g( \hat a^\dag\hat\sigma_-+h.c.)$. 
$\hat F^j(t)$ denotes the operator $\hat F^j(0)$ at time $t$ under the evolution of the adjoint master equation,
\begin{equation}
\begin{aligned}
    \frac{d}{dt} \hat{O}= \mathcal{L}^\dag \hat O=&\frac{i}{\hbar} \left[ \hat{H}_{JC}+\hat{H}_{pump}, \hat{O} \right] \\ &+ \kappa {\mathcal{D^{\dagger}}[\hat{a}]\hat{O}} +\Gamma {\mathcal{D^{\dagger}}[\hat{\sigma}_{-}]\hat{O}}.
\end{aligned}
\label{eq: motion of equation}
\end{equation}
Clearly, the diffusion tensor is proportional to $\partial_i g\partial_j g$, regardless of the strength of the pump field $\varepsilon$.
In the atom-cavity system, the momentum diffusion mainly comes from the cavity-assisted optical dipole force and the spontaneous emission into free-space. 
Since directly applying $\hat\rho^{\rm weak}_0$ requires evolving the adjoint master equation, we will follow the procedures in Ref.~\cite{hechenblaikner1998cooling}, and apply the quantum regression theorem to calculate the diffusion coefficient in the weak-driving limit analytically:
\begin{subequations}
\label{eq: diffusion equation}
\begin{align}
    \label{eq: Ddp expression}
    &{\bf D}_{\rm dp} = \hbar^{2} \left( \nabla g \right)^{2} \frac{\varepsilon^{2} \Gamma}{|Q|^2} \left( 1 + \frac{4 \Delta_{a} g^{2}}{\Gamma} \frac{\Delta_{c} \Gamma + \Delta_{a} \kappa}{|Q|^2} \right)\,, \\
    &D_{\rm SE} = \hbar^{2} k^{2}g^{2} \frac{\varepsilon^{2} \Gamma}{|Q|^2}\,,
\end{align}
\end{subequations}
where $\left( \nabla g \right)^{2}$ should be understood as a matrix $\partial_i g\partial_j g$, and ${\bf D}_{\rm SE}=D_{\rm SE}{\bf I}$ is a scalar matrix. 
The full derivation is shown in Appendix~\ref{app: diffusion}. 

Beyond weak-driving, Eqs.~(\ref{eq: diffusion equation}a) and (b) no longer hold, and we need to evaluate $\hat{\rho}_0$ and Eq.~(\ref{eq: diffusion definition}) numerically. 
Specifically, we evolve $\hat F^i(0)\hat{\rho}_0$ according to the master equation instead of $\hat F^j(t)$, as $\langle\hat F^i(0)\hat F^j(t)\rangle=\Tr(\hat F^j(0)e^{\mathcal{L}t}\hat F^i(0)\hat{\rho}_0)^*$.

\begin{figure}[t]
    \centering
    \includegraphics[width=1 \columnwidth]{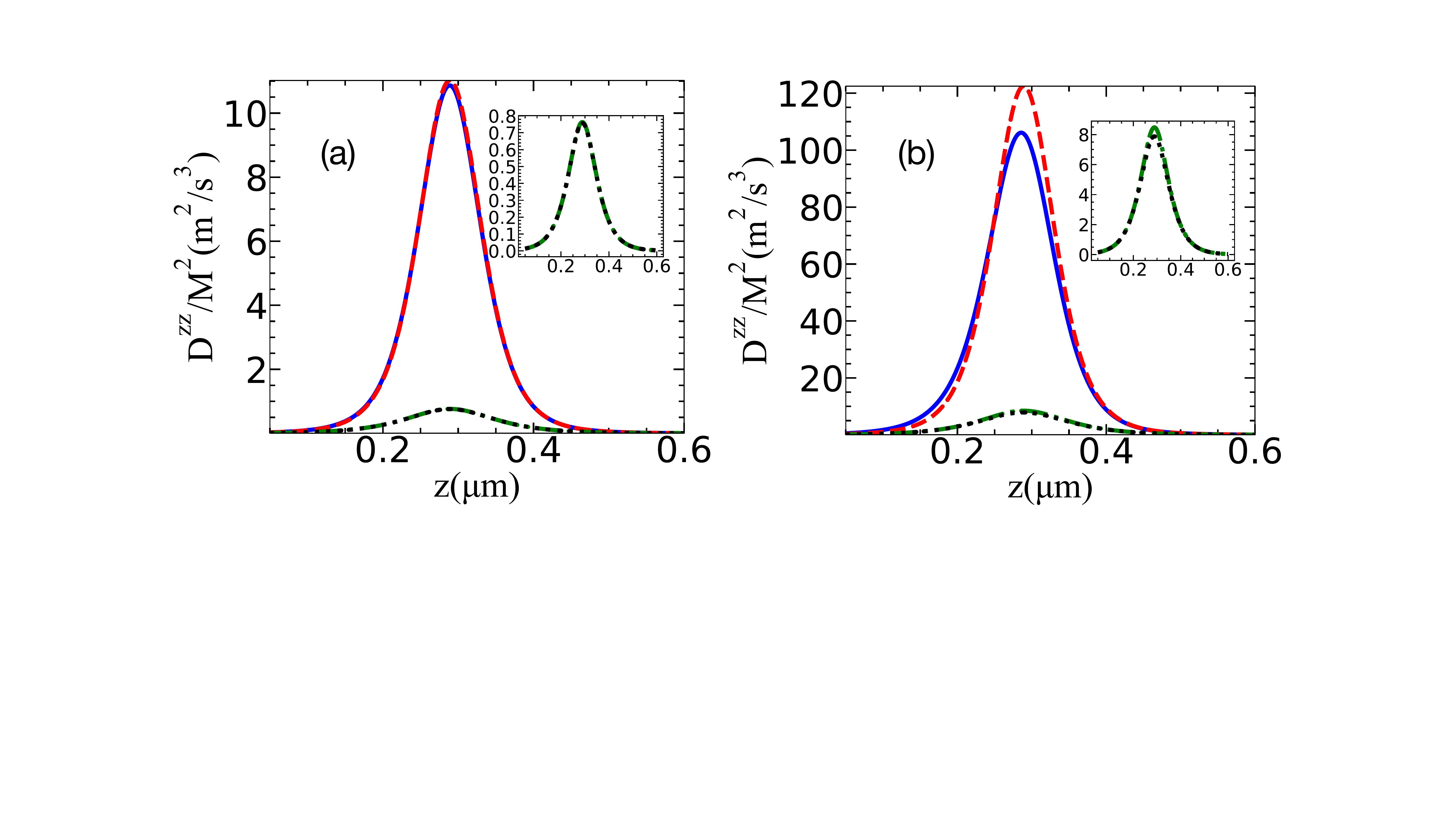}
    \caption{
    Position-dependent diffusion coefficients $D^{zz}(0,0,z)$ with $\varepsilon/2\pi = $(a) 3, and (b) 10~MHz, respectively. 
    Solid (dashed) curves display the contribution from dipole force $D_{\rm dp}$ calculated numerically (analytically). 
    Dotted (dash-dotted) curves show the numerical (analytical) results for the contribution from spontaneous emission $D_{\rm SE}$; insets display a smaller scale.
    }
    \label{fig: diffusion coefficient}
\end{figure}

Similar to the friction coefficient $\boldsymbol{\beta}$, 
we focus on the dynamics of momentum diffusion ($D^{zz}$) along the $z$-direction. 
In Fig.~\ref{fig: diffusion coefficient}, two coefficient components, $D_{\rm dp}^{{zz}}$ and $D_{\rm SE}^{{zz}}$, evaluated from two methods are displayed. 
We note that both diffusion coefficients peak at the same location as in $\expval{F_z}$ in weak-driving.
$D_{\rm dp}^{{zz}}$ is more than ten-fold larger than $D_{\rm SE}^{{zz}}$ in the region of interest, dominating momentum diffusion. 
This is because of a cavity-enhanced fluctuation and fast evanescence decay, $|\partial_z g /g| = d^{-1} \approx 10/\mu$m$ > k$. 
Therefore, we ignore $D_{\rm SE}^{{zz}}$ in the following discussion but will include it in the full Monte-Carlo simulation. 

The magnitude of the diffusion coefficient plays an important role in loading atoms into a conservative trap. 
According to Eq.~(\ref{eq: diffusion equation}), the magnitude of $D_{\rm dp}$ is proportional to the pump intensity squared; $D_{\rm dp}$ increases by one order of magnitude when $\varepsilon/2\pi$ rises from 3~MHz to 10~MHz. 
However, due to the saturation effect, $D_{\rm dp}$ could not scale up indefinitely. 
At a $10~$MHz pump rate and near $z\approx 300~$nm, we see that velocity diffusion in a time interval of $\Delta t\sim 100~$ns is $\sqrt{D_{\rm dp}\Delta t}/M \approx 3~$mm/s. This suggests velocity diffusion can account for a significant fraction of initial velocity, providing a stochastic mechanism for removing kinetic energy while an atom travels through a conservative surface potential. 
This stochastic force allows initial trap loading, followed by slow cooling through damping friction, as we will demonstrate in Sec.~\ref{sec: MC simulation}.

\section{\label{sec: cooling}Nanophotonic cavity cooling of a single atom}

In this section, we discuss how to cool and trap single atoms efficiently in the vicinity of a waveguide. 
As shown in Fig.~\ref{fig: cavity pump potential comparison}, blue-detuned cavity pumping induces a repulsive steady-state force along the $z$-direction on the mid-plane of the waveguide. 
To override this potential barrier and create a stable trap, we introduce a two-color evanescent field trap~\cite{le2004atom, vetsch2010optical, chang2019microring} to provide a tight spatial confinement. 
An atom, once slowed down stochastically, is expected to oscillate in the cooling zone created by the cavity pump field. 

In Figs.~\ref{fig: friction comparison} and \ref{fig: diffusion coefficient}, we find the maximum damping and diffusion coefficients along $z$-direction at around $z = 250~$nm and $z = 290~$nm in the weak-driving regime, respectively. 
This offset is a desired feature, as an approaching free-space atom could first experience larger random force for stochastic cooling, then with larger damping friction for continuous cooling into the trap. 
To prevent an atom from being randomly heated out of the trap by momentum diffusion, we shall set the two-color trap center at a smaller $z$, closer to the maximum damping region and further away from the region with significant stochastic force (large diffusion coefficient).

\subsection{Equilibrium temperature}

We first calculate the local equilibrium temperature from momentum diffusion and damping in the cooling region. 
For axial cooling in a weakly driven Fabry-Perot cavity~\cite{hechenblaikner1998cooling}, the equilibrium temperature of an atom is estimated to be  $k_{B} T_{\rm eq} = -\Bar{D}/\Bar{\beta}$, 
where $\Bar{D}$ and $\Bar{\beta}$ denote the friction and diffusion coefficients averaged within a wavelength~\cite{van2001cooling}.
In 3D, the local equilibrium temperature can be estimated by (Appendix~\ref{app: equilibrium temperature}) 
\begin{equation}
    k_{B} T_{\rm eq} = -\frac1{3}\Tr({\bf D}\boldsymbol{\beta}^{-1}).
\label{eq: equilibrium temperature}
\end{equation}
We comment that, in the weak-driving limit, the equilibrium temperature is independent of the pump intensity and the mode field profile it excites. 
According to Eqs.~(\ref{eq: diffusion equation}) and (\ref{eq: friction calculation}), both ${\bf D}$ and $\boldsymbol{\beta}$ are proportional to the cavity pump intensity $\varepsilon^2$. They also share the same matrix form $\partial_i g\partial_j g$, when ${\bf D}_{\rm SE}$ is negligible compared with ${\bf D}_{\rm dp}$. 
Thus, the equilibrium temperature is primarily determined by the cavity and atomic linewidths, pump detuning, and the atom-photon coupling strength. 

\begin{figure}
    \centering
    \includegraphics[width=.9 \columnwidth]{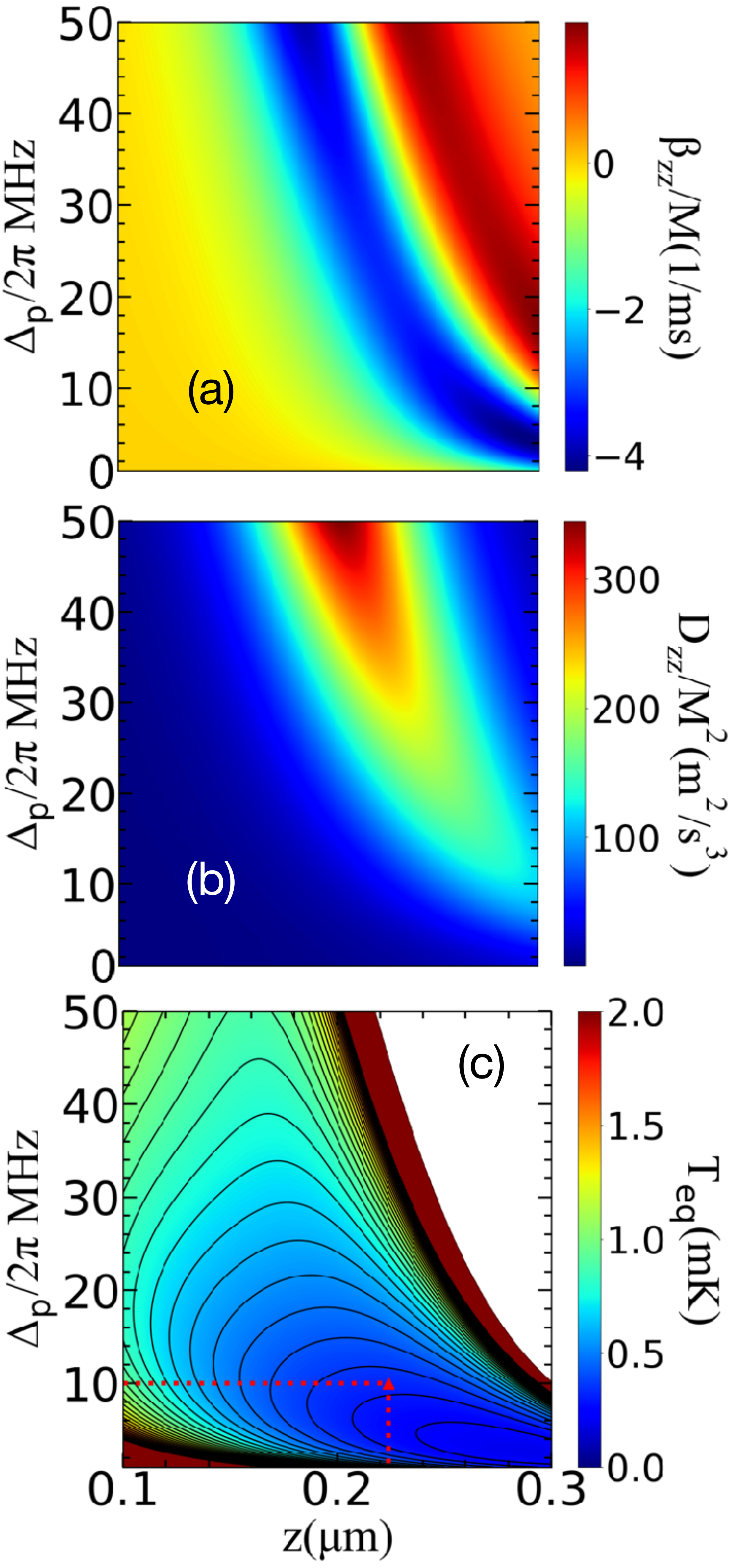}
    \caption{
    (a) Friction coefficients and (b) diffusion coefficients versus pump detuning $\Delta_p$ and atom position $z$. 
    (c) Local equilibrium temperature $T_{\rm eq}(\Delta_p,z)$, estimated using Eq.~(\ref{eq: equilibrium temperature}).
    The triangle symbol and dotted lines denote the selected parameter in Sec.~III~A.
    Other parameters are the same as Fig.~\ref{fig: cavity pump potential comparison}. 
    $\varepsilon=2\pi\times 10~$MHz for (a-b) and (c) is independent of $\varepsilon$.
    }
    \label{fig: equilibrium temperature 2D}
\end{figure}

To show an optimization for equilibrium temperature with the most relevant system parameters, in Fig.~\ref{fig: equilibrium temperature 2D}, we conduct a 2D parameter scan with variable pump detuning $\Delta_{\rm p}$ and atom position $z$ (which controls $g(z)$) while using a moderate pump intensity $\varepsilon=2\pi\times10~$MHz with fixed $\kappa$ and $\Gamma$ as listed in Table~\ref{tab: MC_parameters}. 
We choose this pump rate to achieve sufficiently large cooling efficiency (see Sec. \ref{sec: MC simulation}) while staying close to weak-driving as shown in Figs.~\ref{fig: friction comparison}(c) and Fig.~\ref{fig: diffusion coefficient}(b). 
Overall, given a specific detuning, the $zz$ part of the friction coefficient changes from positive (heating) to negative (cooling) as the atomic position $z$ decreases. 
The maximum cooling and heating positions move toward the waveguide top-surface ($z = 0$) when $\Delta_{p}$ increases in Fig.~\ref{fig: equilibrium temperature 2D}(a). 
This behavior is expected, because upshifting the upper dressed-state energy (to ensure the pump is red-detuned to the dressed-state) requires stronger atom-photon coupling strength near the waveguide surface. 
There are two sweet points to create large damping: one at small $\Delta_{\rm p}$ in the order of $\Gamma$ and at large $z$, and the other one at large $\Delta_{\rm p}$ and small $z$ with strong atom-photon coupling. 
The diffusion coefficient shown in Fig.~\ref{fig: equilibrium temperature 2D}(b), on the other hand, increases with increasing $\Delta_{\rm p}$ and peaks at decreasing $z$ to have a larger contribution of $D_{\rm dp}$ arising from the strong coupling.

As shown in Fig.~\ref{fig: equilibrium temperature 2D}(c), the minimum equilibrium temperature is around 200$~\mu$K if an atom stays trapped near $z \approx 292$~nm. 
However, this is a weak-coupling position with lower cooperativity, which may not be suitable for cavity QED experiments that desire strong atom-photon coupling or high cooperativity. 
Therefore, in the following, we consider a case of larger pump detuning at $\Delta_{p} = 2\pi \times  10~$MHz, with a slightly compromised minimum equilibrium temperature $T_{\rm eq} \approx 324~\mu$K($\equiv T_0$) at $z_{\rm c} \approx 224~$nm. 

We note that, in actual experiments, both an optical trap center and the pump detuning can be tuned initially to achieve higher cooling/loading efficiency at a weak-coupling position. 
Following cavity cooling, the trap center can be adjusted to a strong-coupling position without sacrificing the initial cooling performance.
We also comment that the expected cooling performance is worse than the Doppler limit ($\sim 100~\mu$K) in free-space or in high-finesse cavity cooling. 
This is due to large stochastic force arising from large atom-photon coupling, which is however necessary to slow an atom into a surface trap. 

\begin{figure}[t]
    \centering
    \includegraphics[width=1 \columnwidth]{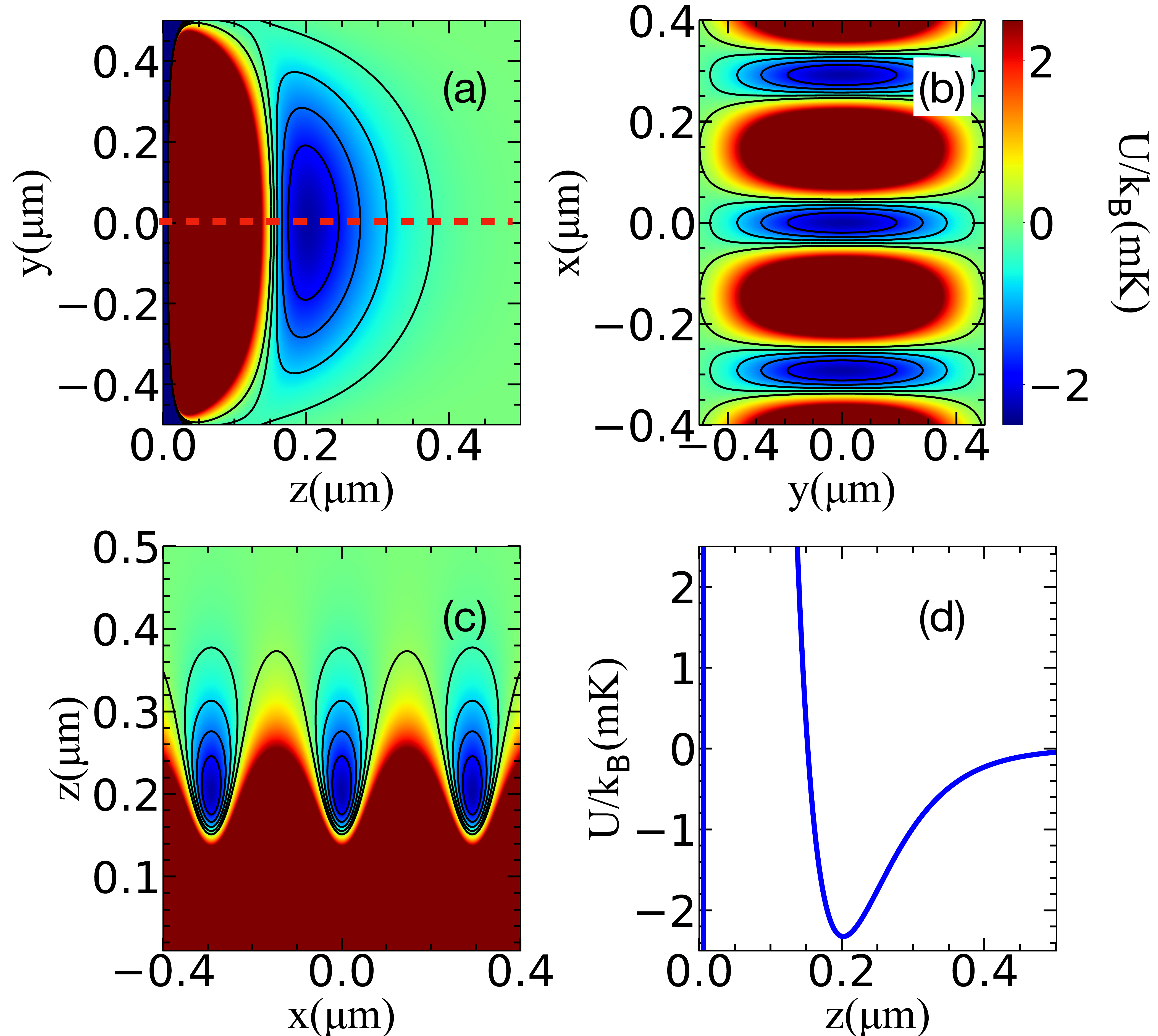}
    \caption{Cross-sections of a two-color evanescent field trap 
    (a) $U(0,y,z)$,
    (b) $U(x,y,z_{\rm t})$, and
    (c) $U(x,0,z)$.
    (d) $U(0,0,z)$ along the dashed line as shown in (a). 
    }
    \label{fig: two-color trap}
\end{figure}

\subsection{Trapping potential}

In this section, we discuss a sample two-color evanescent field trap that localizes an atom in the vicinity of the desired cooling position $z_{\rm c}$. We write the total trap potential in the form
\begin{equation}
    \begin{split}
        U(\textbf{x})=&\alpha_{b}\mathcal{I}_{\rm blue}|\textbf{E}_{b}(\textbf{x})|^{2} + \alpha_{r}\mathcal{I}_{\rm red}|\textbf{E}_{r}(\textbf{x})|^{2}\\
    &-\frac{C_4}{z^3(z+\tilde\lambda)}.
    \end{split}
\end{equation}
where $\alpha_{b(r)}$, $\mathcal{I}_{\rm blue(red)}$, and $\textbf{E}_{b(r)}$ are the scalar polarizability, the intensity, and the normalized electric field profile of the blue-(red-)detuned evanescent field, respectively. 
The last term in the equation is an approximate form of the Casimir-Polder interaction. 
For a cesium atom, we have $C_4/\hbar\approx2\pi \times 267~$Hz$\cdot\mu$m$^4$ and $\tilde\lambda\approx136~$nm to characterize atom-Si$_3$N$_4$ surface interaction~\cite{chang2019microring}. 
Near a rectangular dielectric waveguide, a closed analytical form of electric field does not exist. 
Similar to the cavity mode, we use a single wavevector along each direction to approximate the electric field~\cite{Westerveld2012}. 
Alternatively, the electric fields can be numerically evaluated using finite-element analyses.

An example of two-color trap is plotted in Fig.~\ref{fig: two-color trap}. 
In order to form a tight trap along the waveguide axial ($x$-)direction, we let the blue-detuned field be a traveling wave along the $x$-axis, and the red-detuned field forms a standing wave. 
This can be realized by selectively injecting light from either one or both ends of the waveguide (the cavity Bragg mirrors are assumed to be transparent to these modes with far-off resonant frequencies). 
The two-color trap thus forms a 1D lattice of surface traps along $x$.
Along the vertical $z$-direction, the evanescent fields decay exponentially, but the summation of two-color potentials with two decay lengths forms a stable trap minimum. 
The $z$-location (and depth) of the two-color trap can be adjusted by the relative (and absolute) strength of the blue- and red-detuned fields. 
In Fig.~\ref{fig: two-color trap}, we choose $\mathcal{I}_{\rm red}/\mathcal{I}_{\rm blue}=0.33$ such that the trap center is located at $z_{\rm t}\approx 200$~nm, which is slightly smaller than $z_{\rm c}$. 
This choice takes into account the asymmetric profile of the trap [Fig.~\ref{fig: two-color trap}(d)] and the extended range of low equilibrium temperature shown in Fig.~\ref{fig: equilibrium temperature 2D}(c).

\subsection{Cooling in other directions}
In the previous sections, we mainly focus on the dynamics along the $z$-direction. 
In this section, we investigate the friction and diffusion coefficients around the trap center $z_{\rm t} = 200~$nm and along the $x$ and $y$ directions, seeing whether the cooling condition holds. 
Since ${\bf D}$ and $\boldsymbol{\beta}$ both are $3\times3$ matrices with 3D position dependence (see Appendix~\ref{app: Other direction}), without losing our main focus, we start by calculating the 3D equilibrium temperature based on Eq.~(\ref{eq: equilibrium temperature}).  

\begin{figure}[hb]
    \centering
    \includegraphics[width=1 \columnwidth]{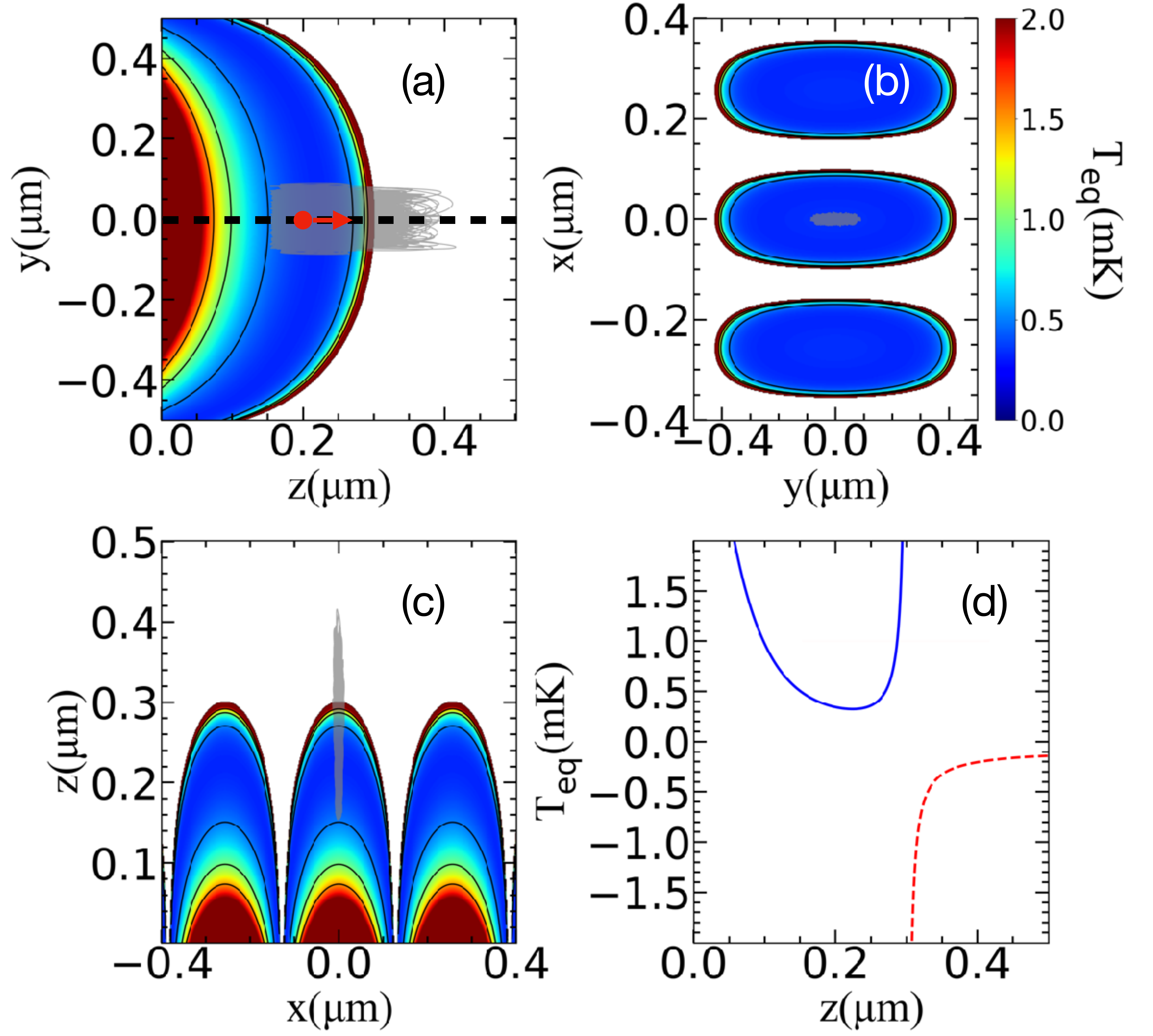}
    \caption{
    Cross-sections of the local equilibrium temperature.
    (a) $T_{\rm eq}(0,y,z)$,
    (b) $T_{\rm eq}(x,y,z_{\rm t})$, and
    (c) $T_{\rm eq}(x,0,z)$.
    (d) $T_{\rm eq}(0,0,z)$ along the dashed line as shown in (a). 
    A negative $T_{\rm eq}$ indicates a heating friction force.
    Shaded regions in (a-c) are projections of a single trapped atom trajectory ${\bf x}(t)$ defined in Eq.~(\ref{eq: MC equations}) that is initially located near the trap center with an initial velocity of 45~cm/s as depicted by the red sphere and arrow in (a).
    }
    \label{fig: Teq3D}
\end{figure}

Using the calculated ${\bf D}$ and $\boldsymbol{\beta}$, we plot the corresponding $T_{\rm eq}({\bf x})$ with sample cross-sections in the $y-z$, $x-y$, and $x-z$ planes, intersecting at the trap center ${\bf x}_{\rm t}=(0,0,z_t)$. 
The result is shown in Figs.~\ref{fig: Teq3D}(a-c). 
We find a broad 3D cooling range (color-shaded regions) nearly covering the entire single-site trap region. 
Since the coupling constant $g$ is periodic, the cooling range also appears periodically along the cavity axis, where we have assumed the anti-node of the cavity pump field overlaps with the center of a single site in the two-color trap at $x=0$. 
We note that, due to axial wavenumber mismatch in the pump and the red-detuned trap fields, some trap sites at large $|x|$ may shift into the heating zone (white region).   

Comparing Fig.~\ref{fig: Teq3D} with Fig.~\ref{fig: two-color trap}, in the $x-y$ and $y-z$ planes we find the central trap region primarily overlaps with the cooling zone within $T_{\rm eq}\lesssim 0.5~$mK. 
Trapped atomic motion is expected to oscillate within the corresponding region bounded by $U-U_0\lesssim 0.5~$mK, where $U_0$ denotes the local potential minimum. 
Figure~\ref{fig: Teq3D}(a-c) plots the projection of simulated trajectories, illustrating the motion of cavity-cooled atoms near the trap center. 
The semi-classical trajectory of atoms is simulated by the Monte-Carlo method, which we will discuss in detail in the next section. 

We now comment on the role of individual components in $\boldsymbol{\beta}$ and ${\bf D}$ matrices.  
Inside the cooling zones, we find the friction coefficient $\beta_{xx}<0$ is negative but one order of magnitude smaller than $\beta_{zz}$. 
Similar conclusion can be made for $\beta_{yy}$ (and $D_{xx,yy}$ as well), making $\beta_{zz}$ (and $D_{zz}$) the dominant components for cooling. 
This can be understood as $\beta_{ij}, D_{ij}\propto \partial_i g\partial_jg$ and $\partial_x g\approx \partial_y g \approx 0$ near the trap center. 
We point out that $D_{xx}$ and $\beta_{xx}$ do increase drastically beyond the cooling zone along the $x$-axis. 
This, however, does not affect cooling as long as the two-color trap provides tight enough confinement to prevent an atom from wandering into the heating zone. 
Lastly, the off-diagonal terms provide kinetic energy mixing along different directions toward 3D thermal equilibrium. 

\section{\label{sec: MC simulation} Monte-Carlo Simulation of Cavity Cooling}
In our semiclassical model in Sec.~\ref{sec: model}, the atomic center of mass motion is treated classically. 
This approximation holds in general because the two-color trap frequency is $\sim O(100)$~kHz~\cite{chang2019microring}, which is more than an order of magnitude smaller than the atom decay rate. 
To model the cooling performance, we perform 3D Monte-Carlo simulations by numerically solving the Langevin equations,
\begin{subequations}
\label{eq: MC equations}
\begin{align}
\frac{dx^i}{dt} & = v^i,\\
M \frac{dv^i}{dt} & = -\partial_i U+\langle \hat F^i\rangle - \beta^{ij}v_j+B^{ij}w_j,
\end{align}
\end{subequations}
where we have neglected heating effect from the far-off resonant two-color trap. 
We adopt 4$^{\rm th}$-order Runge-Kutta method to numerically solve the non-stochastic part of the differential equation with a time step $\Delta t$. 
For the stochastic part, we directly apply the Euler-Maruyama method, where a Gaussian random vector with variance $\overline{w_i(n\Delta t) w_j(n'\Delta t)}=\frac{1}{\Delta t}\delta_{ij}\delta_{nn'}$ and zero mean is generated for evolution in each step, where we used $\Delta t=8$ns in numerics. 
Here, $\overline{\cdot}$ denotes stochastic averaging, and $\delta_{ij}$ is the Kronecker delta. 
The Gaussian random vector is multiplied by the $B^{ij}$ matrix to account for the stochastic force applied during the time interval $(n,n+1)\Delta t$.

We use the parameters listed in Table~\ref{tab: MC_parameters} to construct an atom-cavity system and a two-color trap.
In the following, we conduct 3D Monte-Carlo simulations with these parameters unless otherwise specified. 
We monitor atomic trajectories within a single trap without considering the periodic lattice potential. 
An atom is considered lost when escaping from the potential well along any direction. 

\subsection{Cavity cooling of an atom inside the trap}

\begin{figure}[t]
    \centering
    \includegraphics[width=1 \columnwidth]{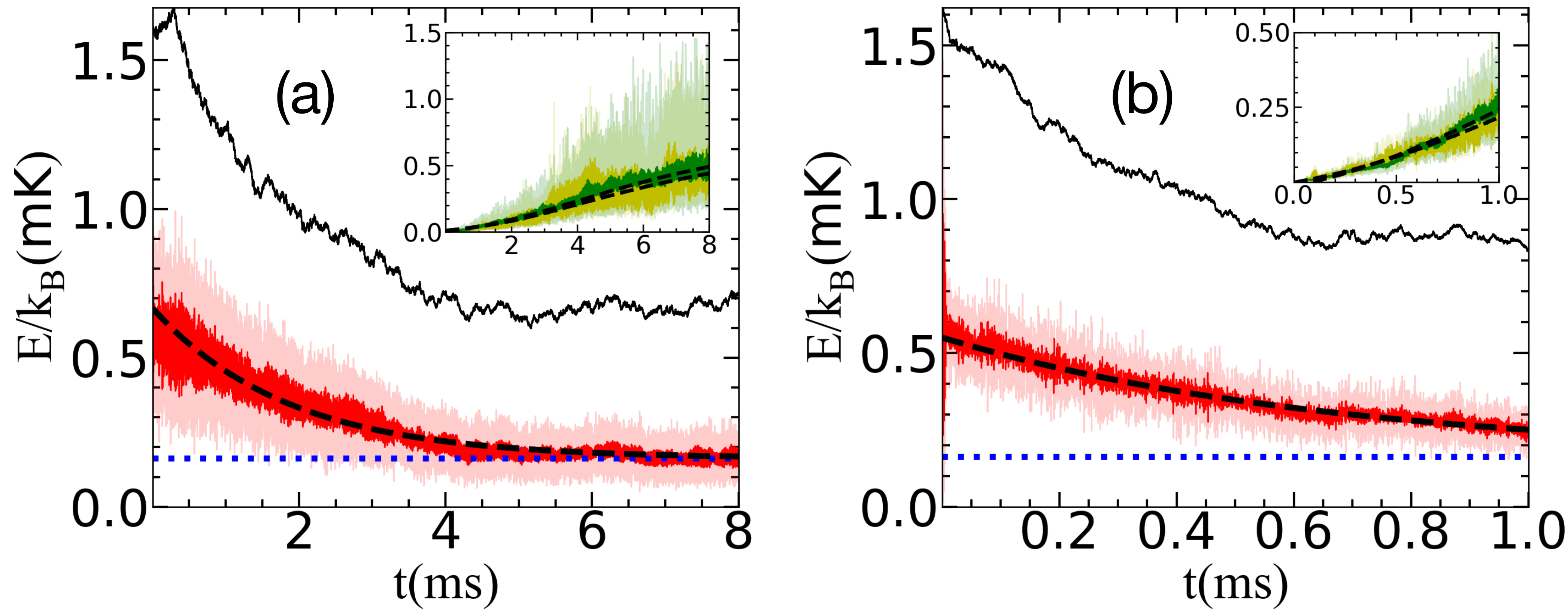}
    \caption{
    Averaged total energy (black curves) and kinetic energy (light red curves) along $z$-direction, obtained from Monte-Carlo simulations of an atom with an initial position at $z_t$ = 200~nm and initial velocity of 45~cm/s moving away from the waveguide, for $\varepsilon/2\pi=$ (a) 5MHz and (b) 10MHz, respectively.
    Red curves show the averaged kinetic energies over a moving time-window (0.8 $\mu$s) to reduce fluctuations.
    Insets show $v^2_x/v^2_z$ and $v^2_y/v^2_z$ (light color curves) and their time-moving average (darker color curves). 
    Dashed lines are obtained from exponential fits, and dotted lines denote $\frac1{2}k_B T_0$, where $T_0=324\mu$K is the minimum equilibrium temperature.
    }
    \label{fig:inside_trap}
\end{figure}

First, we simulate the behavior of an atom initially located near the trap center.
To demonstrate the stability of cavity cooling, we initialize an atom at $z_t=200$~nm with an initial velocity $v_{0} = 45~$cm/s smaller than the escape velocity. 
This corresponds to an initial kinetic energy $E_{\rm k}\approx 1.6~$mK. 
The resulting energy, averaged up to $\sim 50$ trajectories, is shown in Fig.~\ref{fig:inside_trap}. 
These initially hot atoms could be efficiently cooled down within 4 and 0.6~ms with $\varepsilon/2\pi= 5~$MHz and 10~MHz, respectively. 
The potential energy, which is not plotted here for simplicity, is slightly greater than the kinetic energy due to an anharmonic correction in the asymmetric trap~\cite{Cini2012}.
The total energy becomes stable at $k_B\times$0.7-0.8~mK, and the kinetic energy in $z$-direction approaches $k_{\rm B}T_0/2$, which is denoted by blue dotted lines in Fig.~\ref{fig:inside_trap}.
We notice that the balanced cooling and heating effects in $z$-direction and $x,y$-directions cause the plateau in total energy. 
According to Appendix~\ref{app: equilibrium temperature}, we fit the averaged kinetic energy along each direction with an exponential function and find effective friction coefficients $\beta_{\rm eff}/M=-(0.08,0.07,0.54)~$ms$^{-1}$ and $-(0.51,0.41,1.48)~$ms$^{-1}$ along the $x,y,z$-directions for $\varepsilon/2\pi=5$ and $10$ MHz, respectively. 
Full 3D equilibrium will happen at a longer time when $x,y$-directions fully thermalize. 
A sample trajectory is shown in Fig.~\ref{fig: Teq3D} (a-c).

We comment that the cooling rate may further increase if we increase $\varepsilon$. 
However, it may not speed up a lot faster due to the saturation effect. 
Moreover, a larger pump intensity increases the stochastic force. 
It will eventually cause an atom to be randomly heated out of the trap since we have a finite trap depth. 
Random escapes can nevertheless be suppressed by increasing the potential depth.
In the following studies, we choose $\varepsilon = 2\pi \times 10~$MHz to save the simulation time while avoiding a large stochastic force. 

\subsection{Loading free-space atom into a trap}

\begin{figure}
    \centering
    \includegraphics[width=1 \columnwidth]{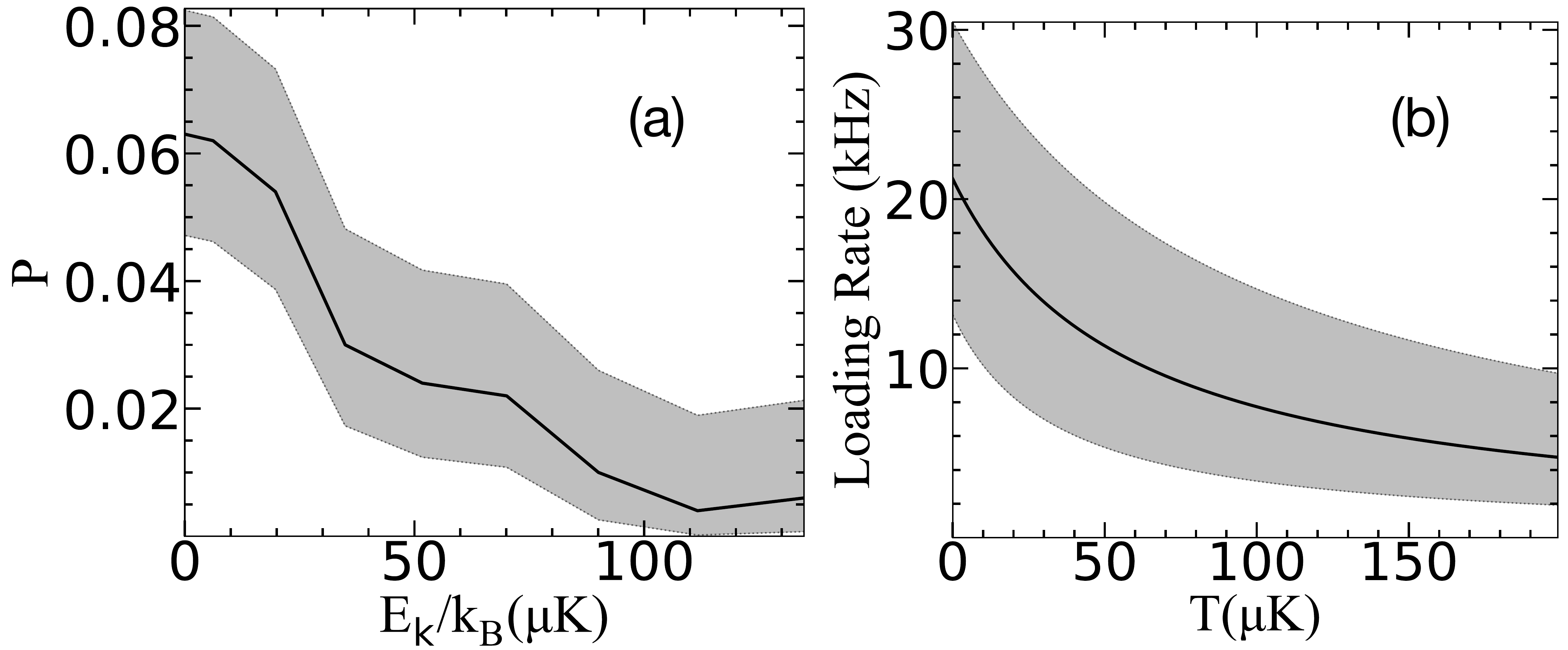}
    \caption{
    (a) Monte-Carlo simulation of trapping probability at $t$=800~$\mu$s as a function of the initial kinetic energy $E_{\rm k}$ at asymptotic infinity. 
    (b) Trap loading rate with an atom flux of 400~kHz prepared at different initial temperature $T$. 
    Shaded regions denote 95\% confidence intervals. 
    }
    \label{fig: side Monte-Carlo}
\end{figure}

Finally, we simulate cavity-assisted free-space atom loading into a two-color trap. 
In our Monte-Carlo simulations, the atoms are initialized at $z_0 = 500~$nm, where the dipole potential is $U(z_0) = -k_B \times 45~\mu$K. 
With an initial velocity of $v_z = -8~$cm/s, which corresponds to initial kinetic energy $E_{\rm k} = k_B \times 6~\mu$K at asymptotic infinity, 31 out of 500 trajectories could be trapped and reach equilibrium after $0.8~$ms of cooling time. 
This projects to $P \sim 6~$\% of trap probability. 
A sweep of trap probability versus initial kinetic energy at asymptotic infinity is shown in Fig.~\ref{fig: side Monte-Carlo}(a). 
When we increase the initial velocity, an atom is less likely to be trapped since the possibility of larger energy dissipation becomes smaller. 
The trap rate decays roughly exponentially, with $P(E_\mathrm{k}) = 0.053 e^{-E_\mathrm{k}/k_B T_{\rm eff}}$, where $E_\mathrm{k}$ is the initial kinetic energy and $T_{\rm eff}\approx 57~\mu$K is an effective threshold temperature determined from the fit; $T<T_{\rm eff}$ can be easily realized by polarization-gradient cooled atoms.

Trap probability is significantly reduced if we decrease the pump rate to $\varepsilon=2\pi\times 5~$MHz, lowering the magnitude of friction and diffusion.
This effect suggests the importance of momentum diffusion for decelerating atoms in a surface potential well. 
In \textit{axial} cavity cooling, an atom can pass through several potential barriers before it rests in one lattice site, reaching thermal equilibrium. 
On the other hand, \textit{transversely} loading an atom into a surface trap is more demanding because there is only one potential barrier above the surface. 
Sufficient cooling must occur in a single pass or within a round trip. 
While friction alone may be insufficient (as in Fig.~\ref{fig: friction comparison}), momentum diffusion creates a stochastic force that is large enough to dissipate the kinetic energy gain from entering the trap. 
Therefore, a larger diffusion coefficient is favored.

In our simulation, we also considered non-zero $v_x, v_y$, spanning finite solid angles. It turns out that only the initial kinetic energy matters as long as an atom could still reach the trap cooling region. 
In a recent experiment~\cite{zhou2023coupling}, a cold atom jet with flux $\sim$400 atom/ms and a small spread of solid angle can be guided toward the waveguide. 
We can estimate a possible cavity-assisted trap loading rate using this technique and under different initial temperature $T$. 
Assuming a constant atom flux, and integrating over the initial velocity under Boltzmann distribution, we obtain a temperature-dependent loading rate, as shown in Fig.~\ref{fig: side Monte-Carlo}(b).
This provides us a loading rate of $\gtrsim$10~atom/ms at an initial temperature $T<50~\mu$K.

\section{Summary}
We investigate the motion of an atom strongly coupled to a weakly-pumped nanophotonic cavity in the presence of a surface micro-trap. 
Analytical and numerical calculations of dipole forces, friction, and diffusion coefficients are reviewed. 
Efficient cavity cooling is considered with a blue-detuned pump field at a sufficiently large intensity marginally beyond the weak-driving limit. 
Using atomic cesium as an example and from full 3D Monte-Carlo simulations, we show that single atoms with an initial temperature $\lesssim50~\mu$K can be efficiently loaded with more than 2\% probability into a milliKelvin deep surface trap. 
Our method provides a viable means to loading single atoms near the surface of a nanophotonic cavity and could potentially be scaled up for loading multiple atoms by considering Doppler cooling using multi-atom dressed states. 

\begin{acknowledgments}
This work was supported by the AFOSR (Grant NO. FA9550-22-1-0031).
\end{acknowledgments}

\appendix
\renewcommand\thefigure{\thesection\arabic{figure}} 
\setcounter{figure}{0} 
\section{\label{app: friction} Expression of Friction Coefficient in the Weak-driving Limit}
In the weak-driving limit, the full expression of the friction coefficient is of the form
\begin{equation}
\begin{aligned}
\beta^{ij} = &-\hbar(\partial_ig\partial_j g) \varepsilon^{2} 4\left(-\Delta_{a}^{3} \Delta_{c}^{4} \Gamma-\Delta_{a}^{2} \Delta_{c}^{3} g^{2} \Gamma+\Delta_{a} \Delta_{c}^{2} g^{4} \Gamma\right. \\
&+\Delta_{c} g^{6} \Gamma-\Delta_{a} \Delta_{c}^{4} \Gamma^{3}+\Delta_{c}^{3} g^{2} \Gamma^{3}-2 \Delta_{a}^{3} \Delta_{c}^{2} g^{2} \kappa \\
&+2 \Delta_{a} g^{6} \kappa+2 \Delta_{c} g^{4} \Gamma^{2} \kappa-2 \Delta_{a}^{3} \Delta_{c}^{2} \Gamma \kappa^{2} \\
&-\Delta_{a}^{2} \Delta_{c} g^{2} \Gamma \kappa^{2}+3 \Delta_{a} g^{4} \Gamma \kappa^{2}-2 \Delta_{a} \Delta_{c}^{2} \Gamma^{3} \kappa^{2}\\
&\left.+\Delta_{c} g^{2} \Gamma^{3} \kappa^{2}-2 \Delta_{a}^{3} g^{2} \kappa^{3} -\Delta_{a}^{3} \Gamma \kappa^{4}-\Delta_{a} \Gamma^{3} \kappa^{4}\right) \frac{1}{|Q|^{6}} \\
&-\hbar(\partial_ig\partial_j g) \varepsilon^{2} 4\left(-2 \Delta_{a}^{2} \Delta_{c}^{2} \Gamma+2 g^{4} \Gamma-2 \Delta_{c}^{2} \Gamma^{3}\right. \\
&-4 \Delta_{a}^{3} \Delta_{c} \kappa+4 \Delta_{a}^{2} g^{2} \kappa-4 \Delta_{a} \Delta_{c} \Gamma^{2} \kappa+4 g^{2} \Gamma^{2} \kappa \\
&\left.+2 \Delta_{a}^{2} \Gamma \kappa^{2}+2 \Gamma^{3} \kappa^{2}\right) \frac{\Delta_a g^{2}}{|Q|^{6}},
\end{aligned}
\label{eq: friction calculation}
\end{equation}
where $Q$ is defined below Eq.~(\ref{eq:density_weak}). 

\section{\label{app: diffusion} Derivation of Diffusion Coefficient in the Weak-driving Limit}
We follow the procedures in the Appendix of Ref.~\cite{hechenblaikner1998cooling} to calculate diffusion coefficient $D$. 
In this section, our goal is to express the formula of $D$ with the expectation value of operators. 
For the spontaneous emission contribution to diffusion coefficient, we quote the result
\begin{equation}
    D_{\rm SE} = \hbar^{2} k_{L}^{2} \Gamma \expval{\hat{\sigma}_{+} \hat{\sigma}_{-}}.
\end{equation}
We note that the geometry of the vacuum mode is different near the waveguide, which may lead to a different $D_{\rm SE}$. 
We here adopt the free space result to compare it with the diffusion coefficient induced by the cavity mode.

When it comes to the contribution of atom-cavity interaction to diffusion coefficients, we first define some vectors and matrices for simplicity:
\begin{equation*}
\vec{Y} = \left(\begin{array}{c}
\hat Y_{1} \\
\hat Y_{2}
\end{array}\right)
=\left(\begin{array}{c}
\hat{a} \\
\hat{\sigma}^{-}
\end{array}\right), 
\quad \mathbf{A}=\left(\begin{array}{cc}
i \Delta_{c}-\kappa & -i g \\
-i g & i \Delta_{a}-\Gamma
\end{array}\right),
\end{equation*}
\begin{equation*}
\vec{X}=\left(\begin{array}{c}
\hat X_{1} \\
\hat X_{2} \\
\hat X_{3} \\
\hat X_{4}
\end{array}\right)=\left(\begin{array}{c}
\hat a^{\dagger} \hat \sigma^{-}+\hat \sigma^{+} \hat a \\
\frac{1}{i}\left(\hat a^{\dagger} \hat \sigma^{-}-\hat \sigma^{+} \hat a\right) \\
\hat a^{\dagger} \hat a \\
\hat \sigma^{+} \hat \sigma^{-}
\end{array}\right),
\end{equation*}
\begin{equation*}
\mathbf{C}=\left(\begin{array}{cccc}
-(\Gamma+\kappa) & -(\Delta_{a} - \Delta_{c}) & 0 & 0 \\
\Delta_{a} - \Delta_{c} & -(\Gamma+\kappa) & -2 g & 2 g \\
0 & g & -2 \kappa & 0 \\
0 & -g & 0 & -2 \Gamma
\end{array}\right), 
\end{equation*}
\begin{equation*}
\quad \vec{I}=\left(\begin{array}{c}
\hat \sigma^{-}+\hat \sigma^{+} \\
-i \left(\hat \sigma^{-}-\hat \sigma^{+}\right) \\
\hat a+\hat a^{\dagger} \\
0
\end{array}\right)
=\left(\begin{array}{c}
\hat Y_{2} + \hat Y_{2}^{\dagger} \\
-i \left(\hat Y_{2} - \hat Y_{2}^{\dagger}\right) \\
\hat Y_{1} + \hat Y_{1}^{\dagger} \\
0
\end{array}\right).
\end{equation*}

From the definition of the diffusion coefficient in Eq.~(\ref{eq: diffusion definition}), the diffusion coefficient arising from atom-cavity interaction is 
\begin{equation}
D^{ij}_{\rm dp}=\hbar^{2}(\partial_i g\partial_j g) \operatorname{Re} \int_{0}^{\infty} d t\left\langle\delta \hat X_{1}(0) \delta \hat X_{1}(t)\right\rangle,
\label{eq: D_sp calculation}
\end{equation}
where an abbreviation is introduced
\begin{equation}
\left\langle a_{\nu}, a_{\mu}\right\rangle=\left\langle\delta a_{\nu} \delta a_{\mu}\right\rangle=\left\langle a_{\nu} a_{\mu}\right\rangle-\left\langle a_{\nu}\right\rangle\left\langle a_{\mu}\right\rangle.
\label{eq: variance abbreviation}
\end{equation}
From the quantum regression theorem and the Heisenberg equations of motion 
\begin{subequations}
\begin{align}
\frac{d}{d t}\left\langle \hat X_{1}(0), \vec{Y}(t)\right\rangle &= \mathbf{A}\left\langle \hat X_{1}(0), \vec{Y}(t)\right\rangle, \\
\frac{d}{d t}\left\langle \hat X_{1}(0), \vec{X}(t)\right\rangle &= \mathbf{C}\left\langle \hat X_{1}(0), \vec{X}(t)\right\rangle+\varepsilon\left\langle \hat X_{1}(0), \vec{I}(t)\right\rangle.
\end{align}
\label{eq: variance motion}
\end{subequations}

A Laplace transform defined for $\xi(t)$ is written as 
\begin{equation}
    L\{ \xi(t) \} = \int_0^{\infty}e^{-st}\xi(t).
\end{equation}
When $s = 0$, considering a Laplace transform 
\begin{equation*}
    L\{\expval{\hat X_{1}(t),\hat X_{1}(0)}\} \equiv L_{\hat X_{1}\hat X_{1}} = \int_{0}^{\infty} d t\left\langle\delta \hat X_{1}(0) \delta \hat X_{1}(t)\right\rangle,
\end{equation*}
 we notice that $L_{\hat X_{1}\hat X_{1}}$ is the key ingredient for calcuating ${\bf D}_{\rm dp}$ in Eq.~(\ref{eq: D_sp calculation}).
 
To solve the value of $L_{\hat X_{1}\hat X_{1}}$, a property of Laplace transform is employed,
\begin{equation}
    L\{ \frac{d}{d t} \xi(t) \} = -\xi(0) + s L\{ \xi(t) \}.
\end{equation}

By applying Laplace transform to both sides of Eqs.~(\ref{eq: variance motion}), following the same denotation we get
\begin{subequations}
\begin{align}
\left\langle \hat X_{1}(0), \vec{Y}(0)\right\rangle &= \mathbf{A} L_{\hat X_{1} \vec{Y}}, \\
\label{eq: Y dagger motion}
\left\langle \hat X_{1}(0), \vec{Y}^{\dagger}(0)\right\rangle &= \mathbf{A}^* L_{\hat X_{1} \vec{Y^{\dagger}}}, \\
\left\langle \hat X_{1}(0), \vec{X}(0)\right\rangle &= \mathbf{C}  L_{\hat X_{1} \vec{X}}+\varepsilon L_{\hat X_{1}\vec{I}}.
\end{align}
\label{eq: Laplace for diffusion}
\end{subequations}
Equation~(\ref{eq: Y dagger motion}) should be taken into account because the components in $\vec{I}$ are a linear combination of in  $\vec{Y}$ and $\vec{Y}^{\dagger}$.

The initial condition $\left\langle \hat X_{1}(0), \hat O(0)\right\rangle$ could be directly calculated from the expectation value of specific operators on steady-states based on Eq.~(\ref{eq: variance abbreviation}). 
In the weak-driving limit, these expressions could be further simplified with quantum noise theory. 
\begin{equation}
\begin{aligned}
\expval{\hat X_{1}(0), \vec{Y}(0)} &= \vec{\mathbf{0}}, \\
\expval{\hat X_{1}(0), \vec{Y^{\dagger}}(0)} &= \left(\begin{array}{c}
\expval{\hat{\sigma}_{+}} \\
\expval{\hat{a}^{\dagger}}
\end{array}\right), \\
\expval{\hat X_{1}(0), \vec{X}(0)} &= \left(\begin{array}{c}
\expval{\hat{\sigma}_{+} \hat{\sigma}_{-}} + \expval{\hat{a}^{\dagger} \hat{a}}\\
-i (\expval{\hat{\sigma}_{+} \hat{\sigma}_{-}} - \expval{\hat{a}^{\dagger} \hat{a}})\\
\expval{\hat{\sigma}_{+} \hat{a}} \\
\expval{\hat{a}^{\dagger} \hat{\sigma}_{-}}
\end{array}\right).
\end{aligned}
\label{eq: initial condition}
\end{equation}

Combining Eqs.~(\ref{eq: Laplace for diffusion}, \ref{eq: initial condition}), we can obtain $L_{\hat X_{1}\vec{X}}$ by solving linear equations and therefore derive the formula of ${\bf D}_{\rm dp}$ in Eq.~(\ref{eq: Ddp expression}).

\section{\label{app: equilibrium temperature} Derivation of local equilibrium temperature}
The stochastic force in the Langevin equation is given by a Wiener process.
For the one-dimensional case, the change in velocity can be written as 
\begin{equation}
  Mdv(t)=\beta v(t)dt+\sqrt{2D}dW(t),
\end{equation} 
where $\overline{dW}=0$ and $\overline{dW(t)dW(t')}=\delta(t-t')dtdt'$.
The final velocity becomes 
\begin{equation}
  v(t)=e^{\beta t/M}v(0)+\sqrt{2D/M^2}\int_0^t e^{\beta(t-s)/M}dW(s).
\end{equation}
Taking the stochastic average, we have 
\begin{equation}
  \begin{split}
    \overline{v^2(t)}=&e^{2\beta t/M}v^2(0)\\+&2D/M^2\int_0^t\int_0^t e^{\beta(t-s')/M} e^{\beta(t-s)/M}\overline{dW(s')dW(s)}\\
    =&e^{2\beta t/M} \bigg(v^2(0)+\frac{D}{M \beta}\bigg)-\frac{D}{M \beta}.
  \end{split}
\end{equation}
The effective temperature is 
\begin{equation}
  k_BT_{\rm eff}=-\frac{D}{\beta}.
\end{equation}
In the three-dimensional case, the velocity increment is
\begin{equation}
  \begin{split}
    M dv_i(t)=&\beta_{ij} v_j(t)dt+B_{ij}dW_j(t),
  \end{split}
\end{equation} 
where $i,j=x,y,z$, repeated indices are summed over.
Suppose the $\beta_{ij}$ matrix is diagonal, we can directly write 
\begin{equation}
  \overline{v^2(t)}=\sum_i\bigg(e^{2\beta_i t/M} \bigg(v_i^2(0)+\sum_j\frac{B_{ij}^2}{2M\beta_i}\bigg)-\sum_j\frac{B_{ij}^2}{2M\beta_i}\bigg)
\end{equation}
In general, when $\boldsymbol{\beta}$ is not diagonal, we can rewrite the equilibrium squared velocity $-\sum\nolimits_{i,j}\frac{B_{ij}^2}{2M\beta_i}$ into $-\Tr({\bf D}\boldsymbol{\beta}^{-1})/M$.
We have used $2{\bf D}={\bf B}{\bf B}^T$.
The equilibrium temperature is 
\begin{equation}
  k_BT_{\rm eff}=-\frac{1}{3}\Tr({\bf D}\boldsymbol{\beta}^{-1}).
\end{equation}

\section{\label{app: Other direction} Friction and diffusion coefficients along other directions}
\begin{figure*}
    \begin{minipage}[t]{0.47\textwidth}
        \centering
        \includegraphics[width=1\columnwidth]{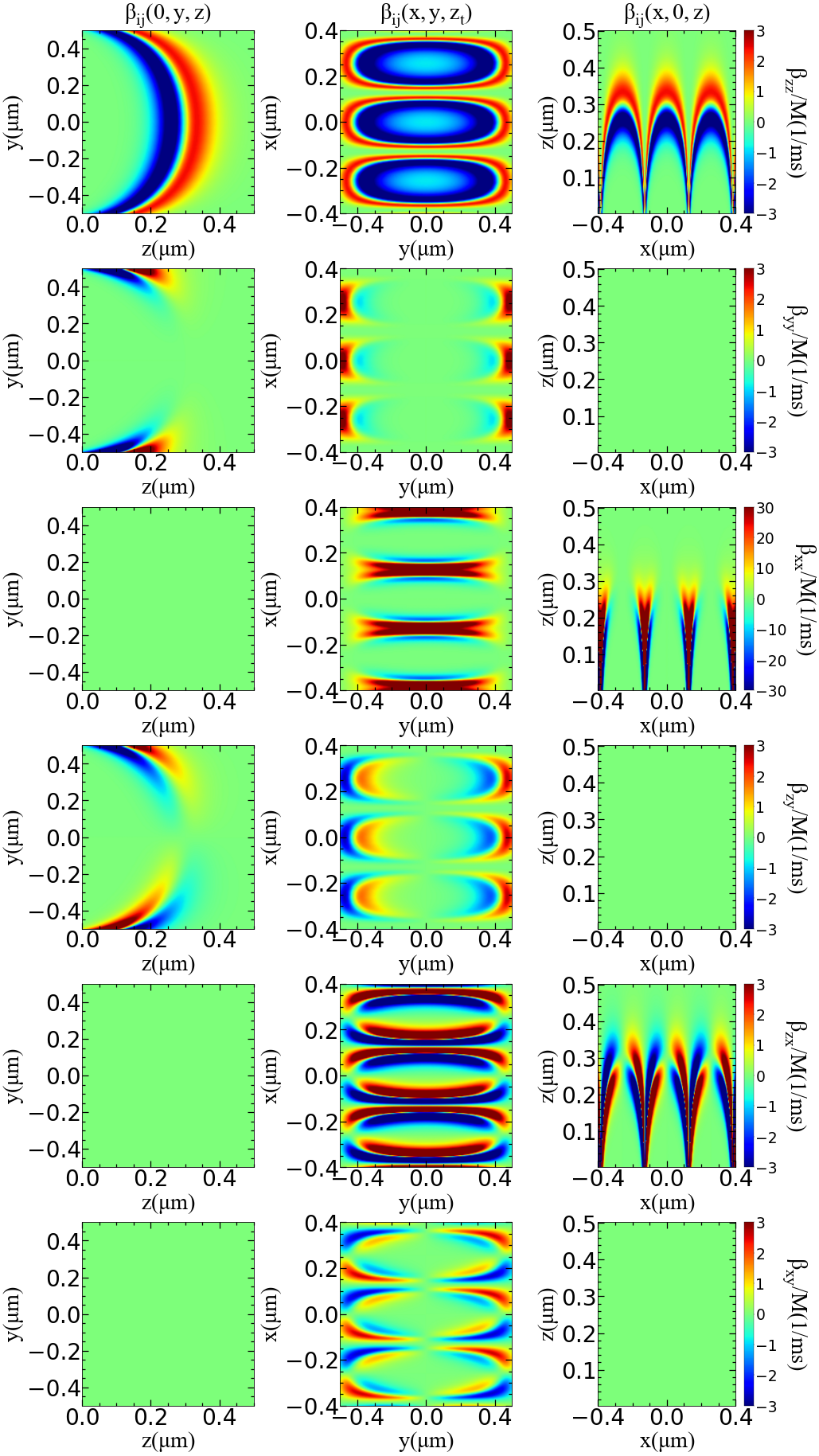}
        \caption{Cross-sections of all different friction tensor  components $\boldsymbol{\beta}$.}
        \label{figs: friction}
    \end{minipage}\hfill
    \begin{minipage}[t]{0.47\textwidth}
        \centering
        \includegraphics[width=1\columnwidth]{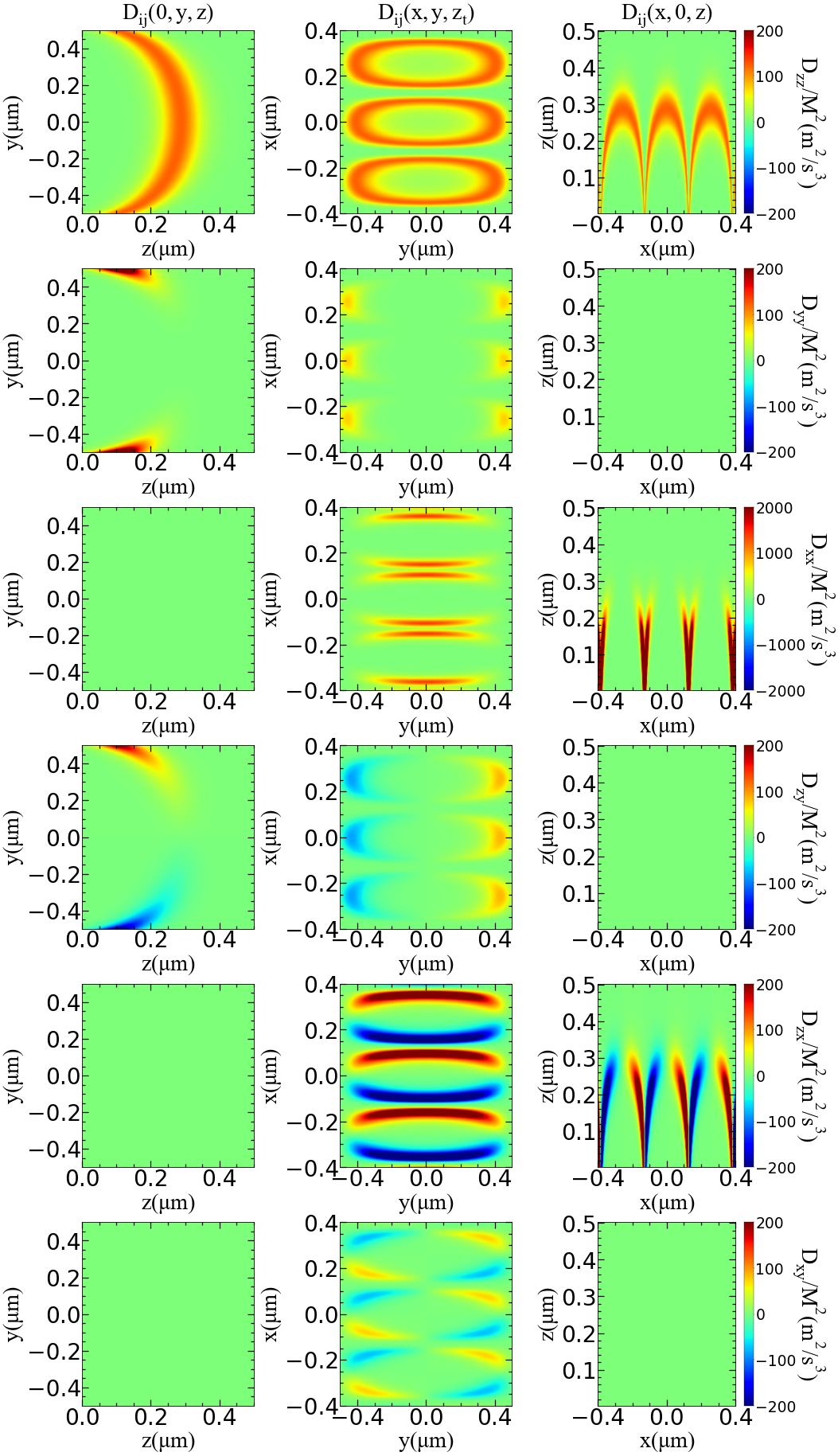}
        \caption{Cross-sections of all different diffusion tensor components ${\bf D}$.}
        \label{figs: diffusion}
    \end{minipage}
\end{figure*}

%
    
\end{document}